\documentclass[twocolumn]{aastex62}
 
\usepackage{graphicx}
\usepackage{subfigure}
\usepackage{multirow}
\usepackage{comment}
\usepackage{natbib}
\usepackage{mathtools}
\usepackage{mathrsfs}
\usepackage{fontenc}
\usepackage{color}
\usepackage{url}
\usepackage{pifont}
\usepackage{textcomp}
\usepackage{booktabs}
\usepackage{array}
\def\loiii{L$_{[\rm O~III]}$}
 
\def\h2{H$_2$} 
 
\def\frii{FR{II}~}
\def\fri{FR{I}}

\def\sii{Si~{\sc ii}}

\def\sii{[S~{\sc ii}]}

\def\o3o2{[O~{\sc iii}]/[O~{\sc ii}]}
\def\o3hb{[O~{\sc iii}]/H$\beta$}
\def\n2ha{[N~{\sc iii}]/H$\alpha$}

\def\oiiib{[O~{\sc iii}]$\lambda$5007~}

\def\oiii{[O~{\sc iii}]$\lambda$5007~}

\def\halpha{H$\alpha$~}
\def\hbeta{H$\beta$~}

\def\kms{km s$^{-1}$} 
\def\apjl{{ApJ}}%
\def\apjs{{ApJS}}%
\def\aap{{A\&A}}%
\begin{document}

\title{X-shaped Radio Galaxies: Optical Properties, Large-scale Environment and Relationship to Radio Structure}
\shorttitle{X-shaped Radio Galaxies}
\shortauthors{Joshi Ravi et al}

\author{ Ravi Joshi}
\affil{Kavli Institute for Astronomy and Astrophysics, Peking University, Beijing 100871, China}

\author{Gopal-Krishna}
\affil{Aryabhatta Research Institute of Observational Sciences ARIES, Manora Peak, Nainital, Uttarakhand 263002, India}
\affil{UM-DAE Centre for Excellence in Basic Sciences, University of Mumbai, Mumbai 400098, India}

\author{Xiaolong Yang}
\affil{Kavli Institute for Astronomy and Astrophysics, Peking University, Beijing 100871, China}
\affil{Shanghai Astronomical Observatory, Key Laboratory of Radio Astronomy, Chinese Academy of Sciences, 200030 Shanghai, P.R. China}

\author{Jingjing Shi}
\affil{Kavli Institute for Astronomy and Astrophysics, Peking University, Beijing 100871, China}

\author{Si-Yue Yu}
\affil{Kavli Institute for Astronomy and Astrophysics, Peking University, Beijing 100871, China}
\affil{Department of Astronomy, School of Physics, Peking University, Beijing 100871, China}

\author{Paul J. Wiita}
\affil{Department of Physics, The College of New Jersey, PO Box 7718, Ewing, NJ 08628-0718, USA}

\author{Luis C. Ho}
\affil{Kavli Institute for Astronomy and Astrophysics, Peking University, Beijing 100871, China}
\affil{Department of Astronomy, School of Physics, Peking University, Beijing 100871, China}

\author{Xue-Bing Wu}
\affil{Kavli Institute for Astronomy and Astrophysics, Peking University, Beijing 100871, China}
\affil{Department of Astronomy, School of Physics, Peking University, Beijing 100871, China}

\author{Tao An}
\affil{Shanghai Astronomical Observatory, Key Laboratory of Radio Astronomy, Chinese Academy of Sciences, 200030 Shanghai, P.R. China}

\author{Ran Wang}
\affil{Kavli Institute for Astronomy and Astrophysics, Peking University, Beijing 100871, China}
\affil{Department of Astronomy, School of Physics, Peking University, Beijing 100871, China}

\author{Smitha Subramanian}
\affil{Indian Institute of Astrophysics, Sarjapur Main Road, Koramangala II Block, Bangalore, Karnatika 560034, India}

\author{Hassen Yesuf}
\affil{Kavli Institute for Astronomy and Astrophysics, Peking University, Beijing 100871, China}

\correspondingauthor{Joshi Ravi}
\email{rvjoshirv@gmail.com (RJ) }

\begin{abstract}
In order to find clues to the origin of the "winged" or “X-shaped”
radio galaxies (XRGs) we investigate here the parent galaxies of a
large sample of 106 XRGs for optical-radio axes alignment,
interstellar medium, black hole mass, and large-scale environment. For
41 of the XRGs it was possible to determine the optical major axis and
the primary radio axis and the strong tendency for the two axes to be
fairly close is confirmed. However, several counter-examples were also
found and these could challenge the widely discussed backflow
diversion model for the origin of the radio wings. Comparison with a
well-defined large sample of normal FR II radio galaxies has revealed
that: (i) XRGs possess slightly less massive central black holes than
the normal radio galaxies (average masses being log$M_{\rm BH} \sim$
8.81 $M_{\odot}$ and 9.07 $M_{\odot}$, respectively); (ii) a much
higher fraction of XRGs ($\sim$ 80\%) exhibits red mid-IR colors ($W2
- W3 > 1.5$), indicating a population of young stars and/or an
enhanced dust mass, probably due to relatively recent galaxy
merger(s). A comparison of the large-scale environment (i.e., within
$\sim$ 1 Mpc) shows that both XRGs and FRII radio galaxies inhabit
similarly poor galaxy clustering environments (medium richness being
8.94 and 11.87, respectively). Overall, the origin of XRGs seems
difficult to reconcile with a single dominant physical mechanism and
competing mechanisms seem prevalent.
\end{abstract}
\keywords{galaxies: evolution, galaxies: kinematics and dynamics,
  galaxies: active, galaxies: star formation, galaxies: radio
  galaxies, galaxies: supermassive black holes}


\section{Introduction}
\label{sec:intro_xrg}
 X-shaped radio galaxies (XRGs) constitute a small but significant
 fraction (up to $\sim$ 3 -- 10\%) of radio galaxies
 \citep{Leahy1984MNRAS.210..929L,Leahy1992ersf.meet..307L,Yang2019arXiv190506356Y}. They
 exhibit two misaligned pairs of radio lobes with the fainter pair
 (called `radio wings') being edge-darkened and the brighter pair
 (called `primary' or `active' lobes) being usually edge-brightened,
 like the classical (FRII) double radio sources
 \citep{Leahy1984MNRAS.210..929L,Capetti2017A&A...601A..81C}. While it
 is generally accepted that the primary lobes are created by a pair of
 powerful radio jets emanating from the central active galactic
 nucleus (AGN), the origin of the secondary pair (wings) continues to
 be debated. \par
The proposed theoretical explanations for the origin of `wings' fall
in two broad categories: (i) intrinsic, i.e., pivoted on the central
engine, and (ii) extrinsic, wherein the external environment plays the
paramount role \citep[see,][for a review]{Gopal2010ApJ...720L.155G}.
According to one popular scenario in the second category, the wings
form due to diversion of the backward flowing synchrotron plasma
within the two radio lobes, as it impinges on an asymmetric
circum-galactic gaseous halo of the parent (early-type) galaxy and is
propelled by the buoyancy forces along the direction of the steepest
pressure gradient in the surrounding medium
\citep{Leahy1984MNRAS.210..929L,Worrall1995ApJ...449...93W,Kraft2005ApJ...622..149K,
  Capetti2002A&A...394...39C,Hodges2010ApJ...717L..37H,Rossi2017A&A...606A..57R}.
For this ``backflow diversion'' model to work, the radio lobes are
required to be of the \frii type, which alone are capable of
sustaining a strong backflow and the absence of hot spots in the wings
is then naturally explained. Strong support to this picture has come
from the optical studies of a small number of XRGs, which have clearly
shown that the direction defined by the pair of radio wings exhibits a
clear preference to be closer to the optical minor axis of the host
elliptical galaxy and the converse holds for the primary lobes
\citep{Capetti2002A&A...394...39C,Saripalli2009ApJ...695..156S,Gillone2016A&A...587A..25G},
Additional support has come from the X-ray studies by
\citet{Hodges2010ApJ...710.1205H} showing that the wings tend to align
with the minor axis of the hot gaseous halo of the parent galaxy.
However, existence of some XRGs with primary lobes of \fri\ type might
challenge this backflow diversion scenario \citep[but
  see,][]{Saripalli2009ApJ...695..156S}.

In the intrinsic class of models, the X-shaped radio morphology is
explained in terms of a rapid change in the jet ejection direction, 
i.e., a reorientation/flip of the spin axis of the supermassive black 
hole (SMBH), either due to merger of a small galaxy with the massive
elliptical host of the radio source
\citep{Zier2001A&A...377...23Z,Merritt2002Sci...297.1310M,Rottmann2001PhDT.......173R,Zier2002A&A...396...91Z},
or due to accretion disk instabilities \citep{Dennett2002MNRAS.330..609D}. 
In the former, so called `spin-flip' scenario, the wings are viewed as 
relics of their pre-merger active phase. Indeed, this model accords well 
with the widely discussed scenario for powerful double radio galaxies,
wherein a galaxy merger of a gas-rich galaxy with the massive elliptical 
triggers the jet activity in the massive elliptical 
\citep{Begelman1984RvMP...56..255B,Wilson1995ApJ...438...62W}. 
Whilst this scenario appears to have found some observational support 
from the detection of SMBH binary within the nuclei of a few AGN
\citep{Rodriguez2006ApJ...646...49R,Kharb2017NatAs...1..727K}, 
and also from the discovery of double-peaked broad emission lines in a 
few AGN \citep{Zhang2007MNRAS.377.1215Z, Rubinur2019MNRAS.484.4933R}, 
it offers no clear explanation for the propensity of the primary radio
lobes in XRGs to be aligned close to the optical major axis of the
parent elliptical galaxy.

Recently, using GMRT observations of 28 XRGs at 610 MHz and 240 MHz,
\citet{Lal2019arXiv190311632L} have found no systematic difference
between the radio spectral indices of the primary radio lobes and the
wings \citep[see also,][]{Lal2005MNRAS.356..232L,Lal2007MNRAS.374.1085L}. 
This appears to be at odds both the `backflow diversion' and the `spin-flip' 
scenarios. To address this, \citet[][]{Lal2005MNRAS.356..232L,Lal2007MNRAS.374.1085L} 
have put forward the `twin-AGN' hypothesis wherein the two pairs of 
lobes are powered by two independent jet pairs emanating from an 
unresolved pair of central engines. However, this picture does not 
explain the preference of the radio wings' pair to be aligned close to 
the optical minor axis of the host elliptical galaxy.

In this context, it is interesting to recall the suggestion that the
jets can be intercepted by rotating segments of stellar/gaseous
shells like  those which have been detected around a number of nearby
early-type galaxies
\citep{Carter1982Natur.295..126C,Schiminovich1994ApJ...423L.101S,
  Oosterloo2005A&A...429..469O,
  Sikkema2007A&A...467.1011S,Struve2010A&A...513A..10S,Mancillas2019arXiv190511356M}.
Earlier, such a scenario of jet-shell interaction has been invoked to
explain certain intriguing radio features witnessed in the nearest
radio galaxy Centaurus A, particularly its peculiar ``North Middle
Lobe'' \citep[] [hereafter GKW10]{Gopal2010ApJ...720L.155G} and the
``North Inner Lobe''
\citep{Gopal1984A&A...141...61G,Gopal1983Natur.303..217G}. Plausibly,
such interactions take place also in XRGs, which would provide a
straight-forward explanation for the strong tendency for their primary
lobes to align with the optical major axis of the host elliptical, as
discussed in
\citet[][, hereafter GBW03, GBGW12]{Gopal2003ApJ...594L.103G,Gopal2012RAA....12..127G}. Such
putative shells have since also been detected in a few (more distant)
elliptical galaxies hosting an XRG, such as 3C 403 and 4C$+$00.58
\citep{Hodges2010ApJ...717L..37H,
  Ramos2011MNRAS.410.1550R,Tadhunter2016A&ARv..24...10T}.

The Z-symmetry of the radio wings in XRGs, highlighted in \citet{Gopal2003ApJ...594L.103G} 
has motivated a hybrid model which reconciles the spin-flip scenario with 
the striking Z-symmetry of the wings and, moreover, can easily explain why 
some wings are longer than the associated primary lobes (see discussion in 
Section~\ref{sec:dis}). In this model, the observed Z-symmetric radio wings 
form via a bending of the twin-jets in opposite directions upon their 
interacting with the ISM which has been set in ordered rotation due to the 
in-spiraliing of a gas-rich galaxy captured by the jet-emitting massive 
elliptical. Later, as the central black holes of the two merging galaxies 
coalesce, the SMBH spin axis (and hence the axis of the emergent jet pair) 
would get aligned towards the initial orbital momentum vector of the captured 
galaxy. Thus, since after the spin-flip, the direction of the twin-jets would 
be close to the ISM rotation axis, their outward propagation to form the primary 
lobe pair, would no longer receive a side-way push from the the rotating ISM,
which would facilitate their linear propagation \citep{Gopal2003ApJ...594L.103G}. 
As argued in GBW03 and GBGW12 \citep[see, also,][]{Hodges2014ApJ...789..131H}, 
such a model can also explain in a natural way the curious result that radio 
luminosities of XRGs tend to cluster near the Fanaroff-Riley transition, i.e. 
$P_{178}$ MHz $10^{25}$ WHz$^{-1}$ sr$^{-1}$.

It needs to be appreciated that the afore-mentioned observational clues are 
still based on rather small samples, namely 11 XRGs found in the 3CR catalog 
\citep{Leahy1992ersf.meet..307L} and 100 XRG candidates identified in a 
systematic search based on the FIRST survey \citep{Cheung2007AJ....133.2097C}. 
Recently, we have expanded the sample by preparing a new catalog of 290 XRG 
candidates, primarily based on of 1.4 GHz FIRST and the TIFR GMRT sky survey 
(TGSS) made at 150 MHz \citep{Yang2019arXiv190506356Y}. This latest catalog of 
XRG candidates is an extension to smaller angular sizes of the XRG search 
undertaken by \citet{Cheung2007AJ....133.2097C}, which led to a catalog of
100 XRG candidates, nearly $\sim 75\%$ of which were confirmed as bona-fide 
XRGs, in follow-up VLA observations \citep{Roberts2018ApJ...852...47R}. In 
our catalog of 290 XRG candidates, 106 were classified as `strong' XRG 
candidates and only this subset will be used in the present study to explore the 
salient scenarios proposed to understand the XRG phenomenon. In particular, we 
shall endeavor to examine the geometrical/morphological relationship between 
the radio lobes and the optical host galaxy, as well as the possible relevance 
of galaxy merger and large-scale environment to the occurrence of XRGs.

\begin{figure*}
\centering
\includegraphics[clip,height=9.5cm,width=8cm]{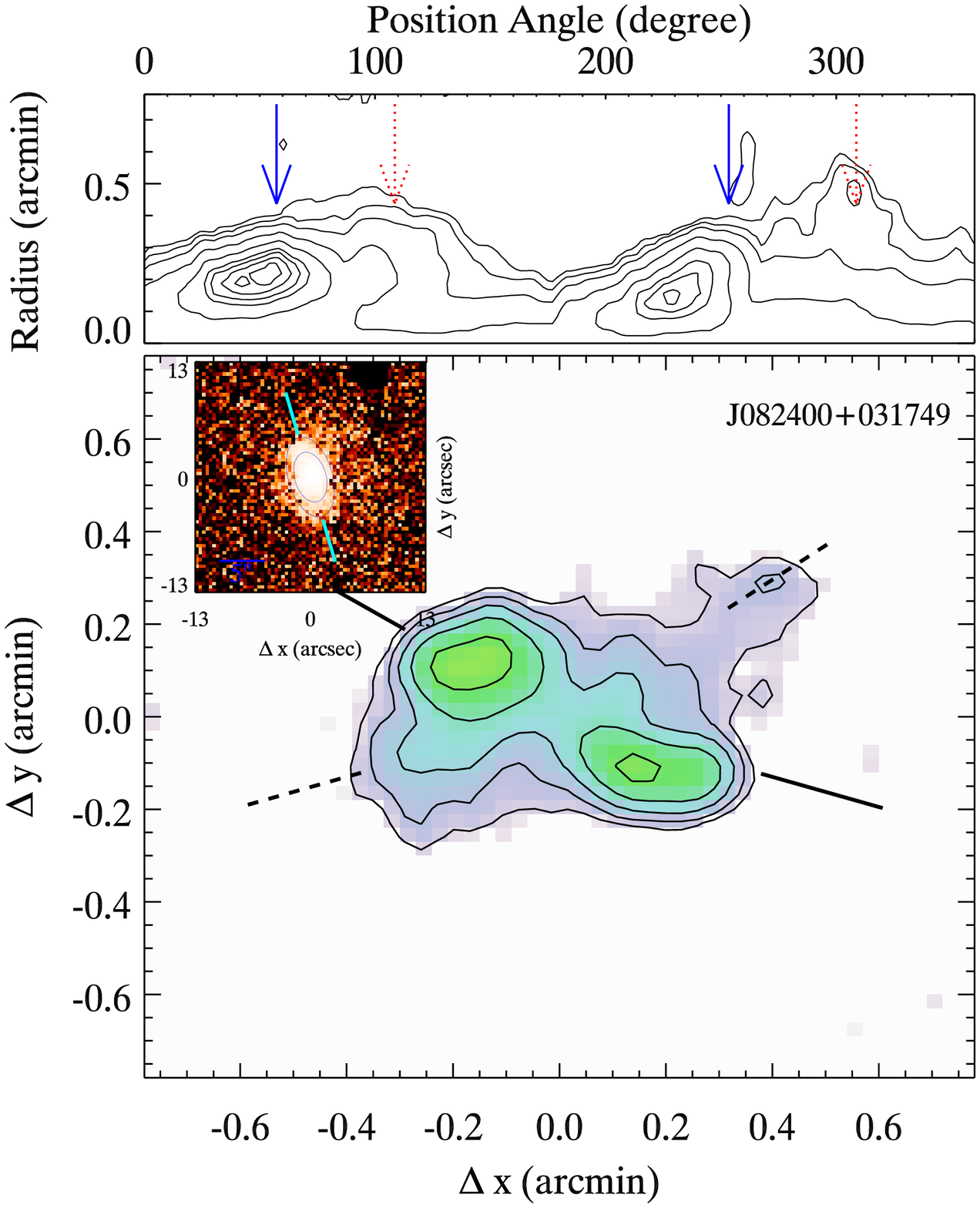}
\includegraphics[clip,height=9.5cm,width=8cm]{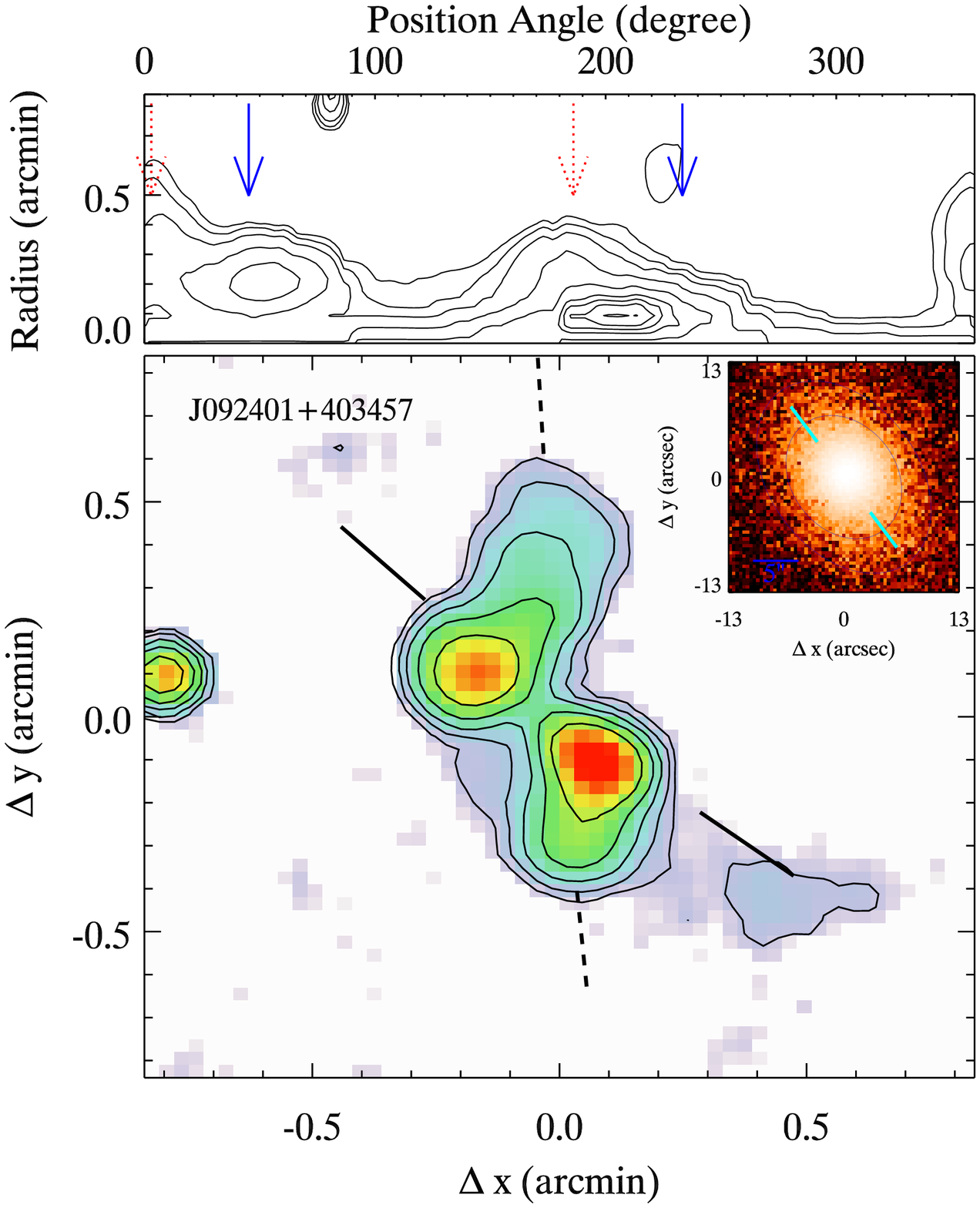}
  \caption{\emph{Lower Left Panel:} The FIRST image of the XRG
     J000450$+$124839 at 1.4 GHz and the corresponding contours
     starting at 3$\sigma$ level. The axes of the radio wings and the
     primary lobes are shown by dotted and solid lines, respectively.
     The inset shows the ($i$-band) SDSS image together with its major
     axis and the fitted ellipses (blue curves). \emph{Upper
       illustration:} Polar diagram of the FIRST radio map at 1.4 GHz.
     The contours start at 3 times the local rms noise level in the
     map (typically at $\sim$0.5 mJy/beam). The two bright peaks (blue
     arrows) mark hot spots in the two primary lobes, whereas the pair
     of radio wings can be identified with the linear extended
     features (red arrows). These identifiers yield quantitative
     estimates of the PAs of the primary and secondary radio axes
     defined by the pairs of the primary lobes and the wings,
     respectively. \emph{Right Panel:} The same for the XRG
     J092401$+$403457.}
 \label{fig:pa_img}
    \end{figure*}

This paper is organized as follows. Section 2 describes our sample of
XRGs, as well as a control sample comprised of normal \frii radio galaxies. 
In Section 3,  we examine the geometrical relationship between radio and 
optical morphologies of XRGs. Section 4 presents a comparison of the XRGs 
and \frii radio galaxies in terms of optical spectral line properties 
and galaxy estimates. The discussion and conclusions of this study are 
summarized in Section 5. Throughout, we have assumed the flat Universe 
with $H_0 = 70$~\kms\ $\rm Mpc^{-1}$, $\Omega_{\rm m} = 0.3$ and 
$\Omega_{\rm \Lambda} = 0.7$.

 \begin{figure}
  \centering
\includegraphics[clip,height=8.2cm,width=8.2cm]{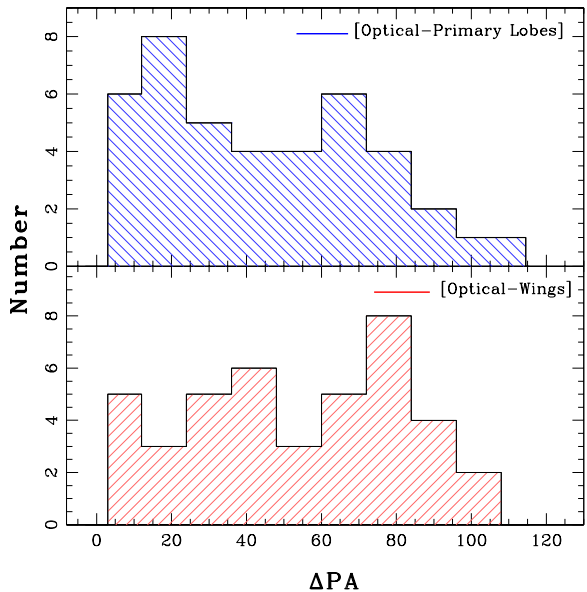} 
  \caption{\emph{Lower Panel:} Histogram of position angle offset of 
 the optical main axis of the host elliptical from the axis defined 
 by the pair of radio wings (secondary lobes). \emph{Upper Panel:} the 
 same for the pair of primary radio lobes.}
\label{fig:pa}
 \end{figure}

\section{The XRG Sample}
\label{sec:sample}
Our starting point is the sample of 106 strong XRG candidates, extracted
from the compilation of 290 winged radio galaxies published recently by
some of us \citep{Yang2019arXiv190506356Y}. In brief, we extended 
the catalog of 100 XRGs with radio major axis, $\theta_{major} \ge 15$
arcsec \citep{Cheung2007AJ....133.2097C} to smaller angular sizes, 
i.e.,  considering sources with $\theta_{major}$  down to 10 arcsec. The resulting
catalog of 290 XRG candidates, mainly based on the 1.4 GHz FIRST and 150 MHz 
TIFR GMRT sky survey (TGSS), includes  106 `strong' XRG candidates in
which at least one of the two wings is well defined (Section~\ref{sec:intro_xrg}).

Given that the active (primary) radio lobes of XRGs are overwhelmingly
of the \frii type and they predominantly occupy the domain of \frii
sources in the Owen-Ledlow Plane (see, Section~\ref{sec:intro_xrg}; \citealt{Yang2019arXiv190506356Y}), 
we are motivated to carry out a comparison of the properties of the 
XRGs with those of \frii radio galaxies. For this, we have used a sample 
of 401 \frii radio galaxies from \citet{Koziel2011MNRAS.415.1013K} for 
which optical spectra are available in the SDSS, except for one object. 
Out of these 400 \frii radio galaxies with spectra, we have excluded ten, as they 
exhibit `double-double' radio morphology \citep{Schoenmakers2000MNRAS.315..371S} 
and another two sources which have been re-classified as XRGs by 
\citet[][]{Kuzmicz2017MNRAS.471.3806K}. These exclusions limit the
control sample to 388 \frii radio galaxies.

 Compared to the \citet{Cheung2007AJ....133.2097C} catalog, 
our source catalog of 290 XRG candidates admits only sources of smaller 
radio (angular) extent and spans a larger redshift range ($0.06 
\le z \le 0.7$, with median $z \sim 0.37$). Hence, for a more general 
comparison with the control sample of \frii radio galaxies which spans 
a wide range in radio power and size, we have expanded our XRG sample 
by including the 11 XRGs reported in \citet{Leahy1992ersf.meet..307L} 
and the 100 XRG candidates cataloged by \citet{Cheung2007AJ....133.2097C}.
We have used information available in the literature for them.

 \begin{figure*}
  \centering
\includegraphics[clip,trim= 5pt 5pt 5pt 5pt,height=5.2cm,width=5.2cm]{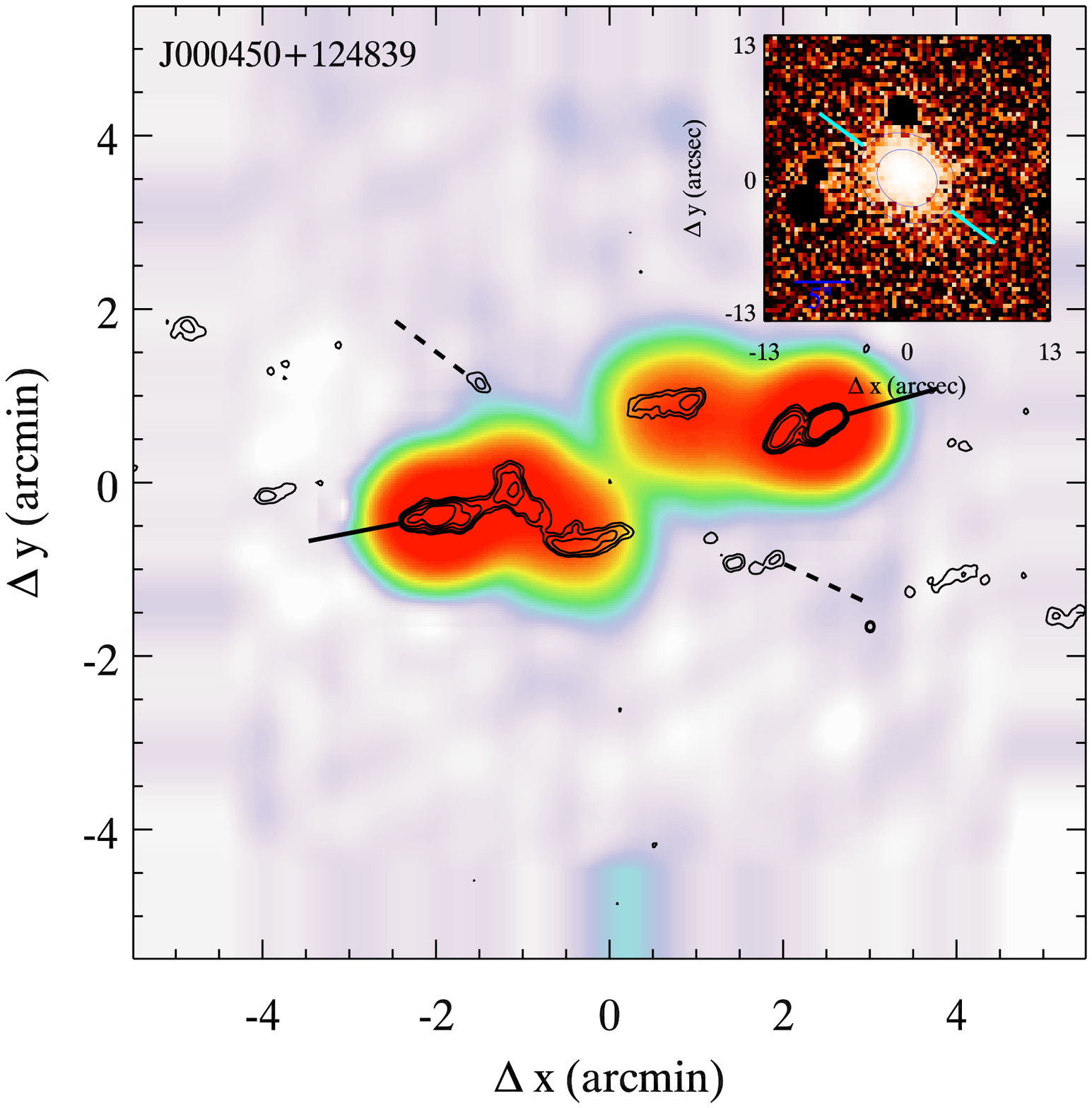} 
\includegraphics[clip,trim= 5pt 5pt 5pt 5pt,height=5.2cm,width=5.2cm]{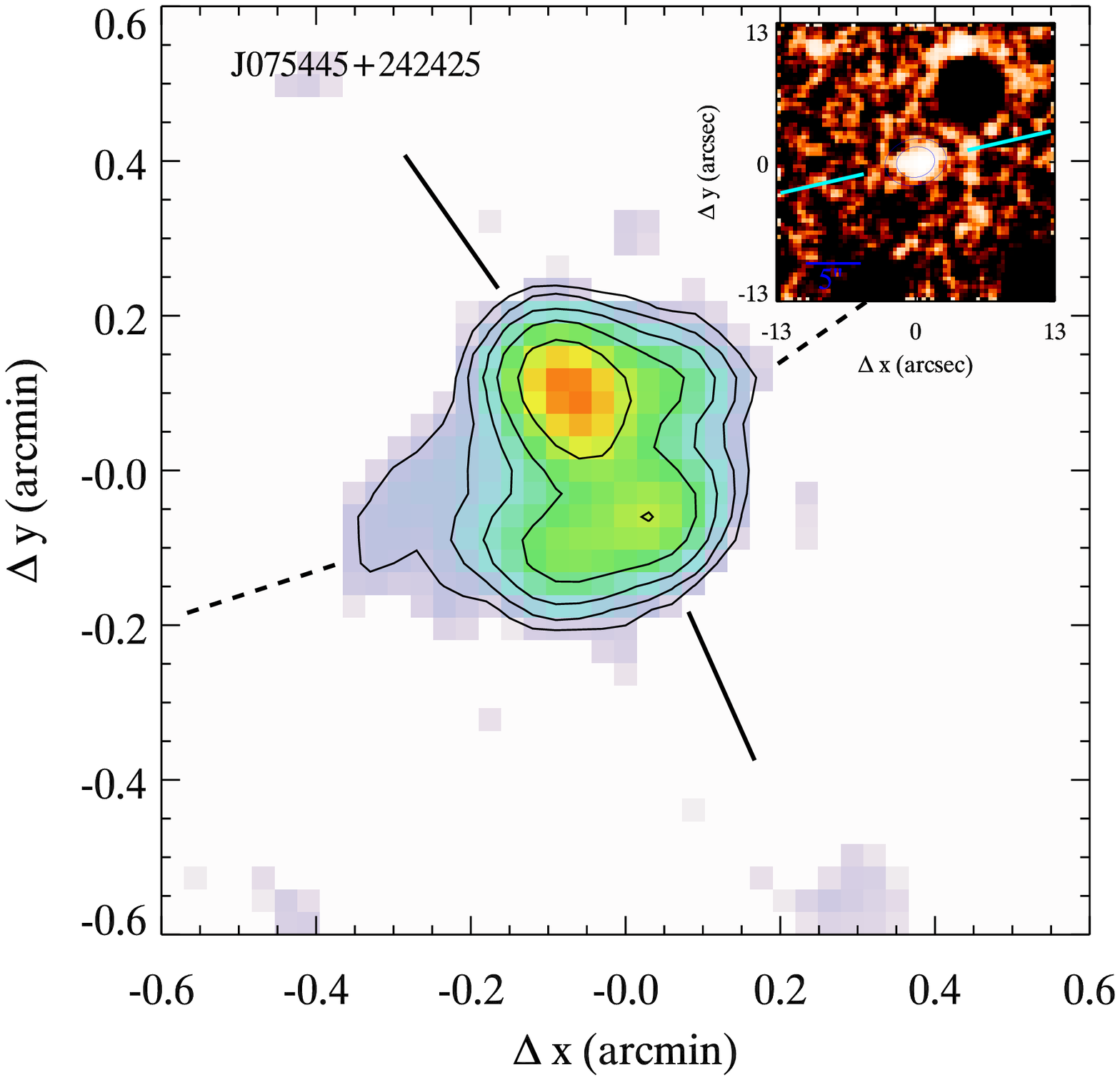} 
\includegraphics[clip,trim= 5pt 5pt 5pt 5pt,height=5.2cm,width=5.2cm]{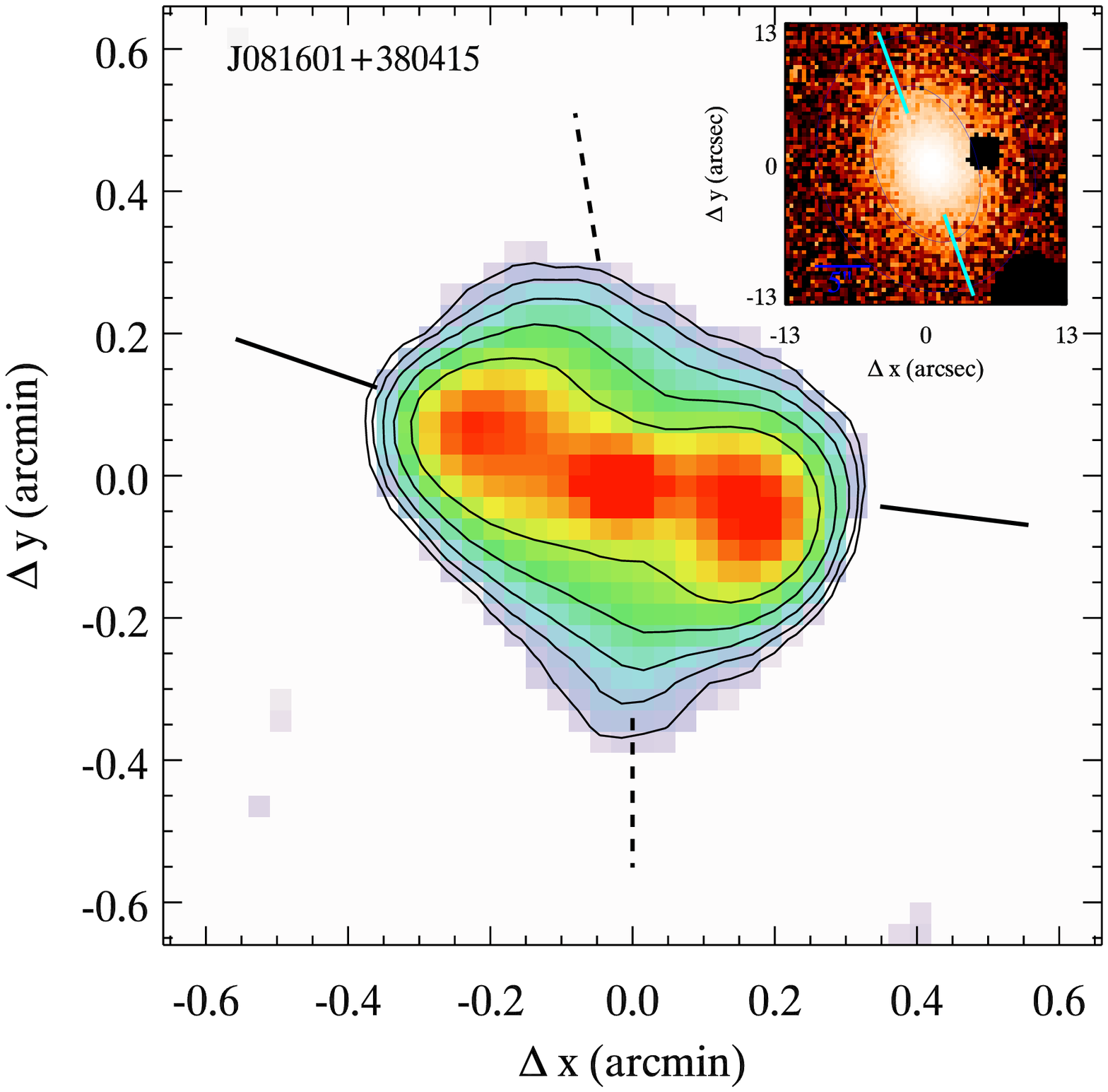} 
\includegraphics[clip,trim= 5pt 5pt 5pt 5pt,height=5.2cm,width=5.2cm]{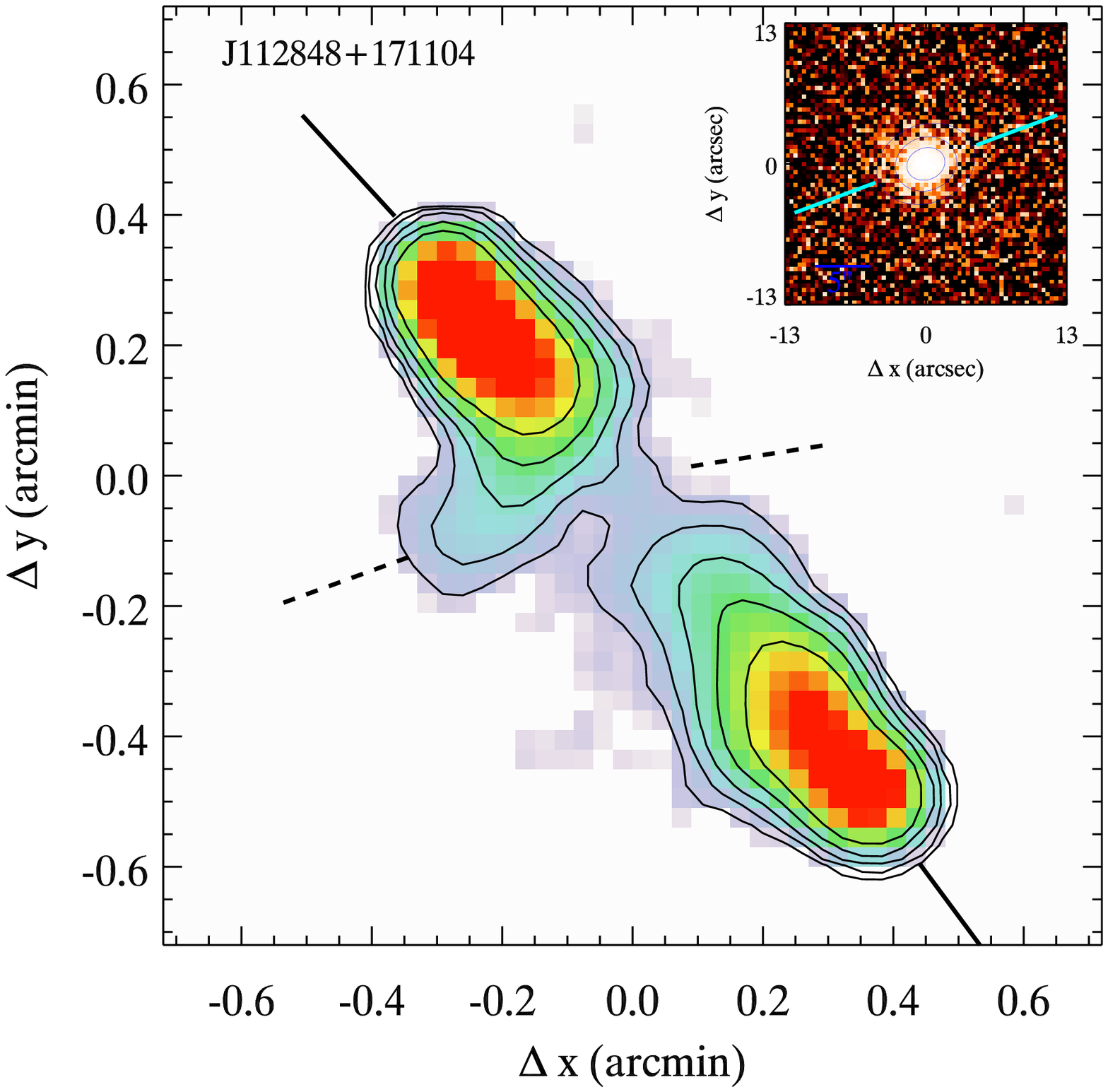} 
\includegraphics[clip,trim= 5pt 5pt 5pt 5pt,height=5.2cm,width=5.2cm]{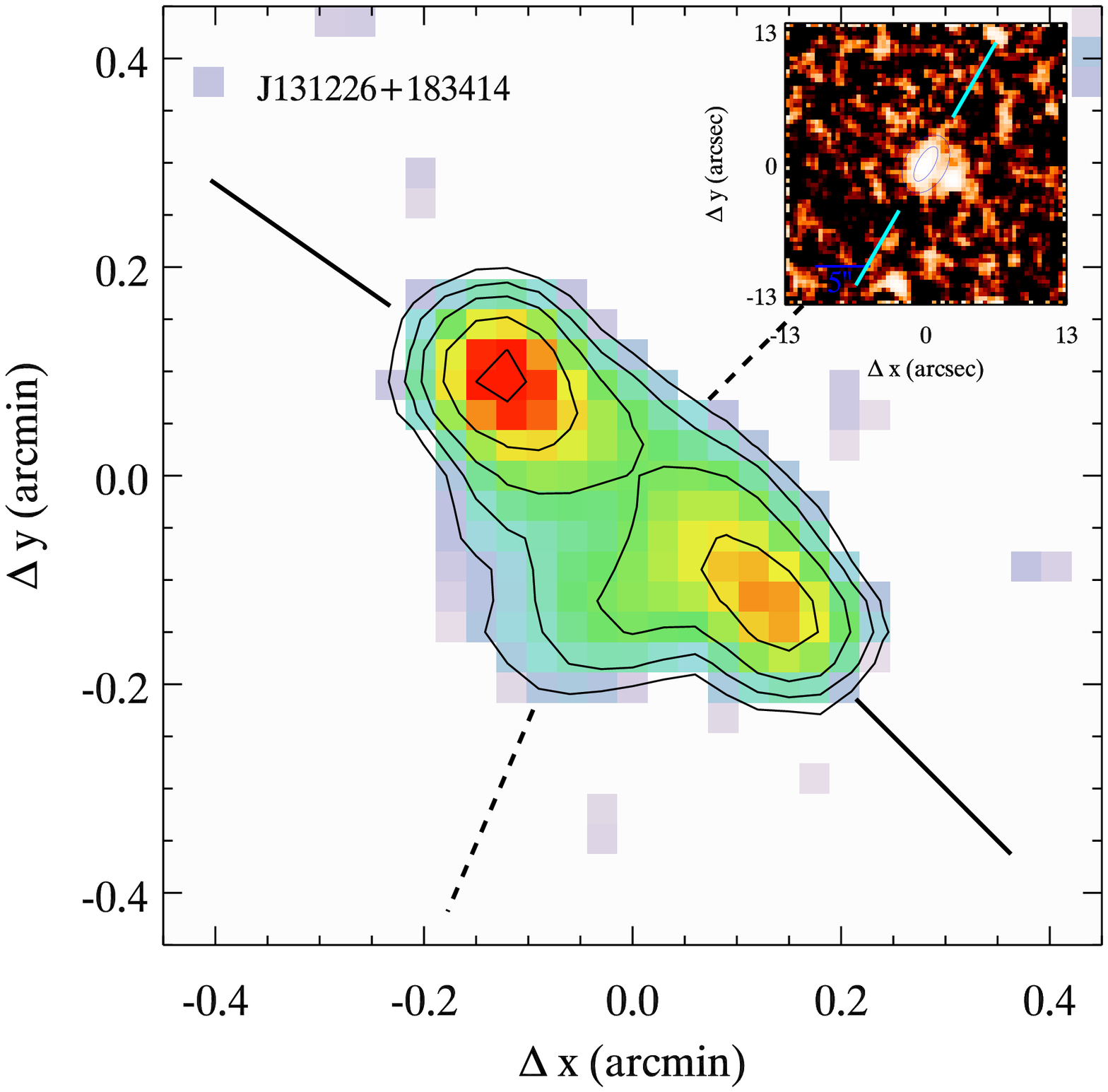} 
\includegraphics[clip,trim= 5pt 5pt 5pt 5pt,height=5.2cm,width=5.2cm]{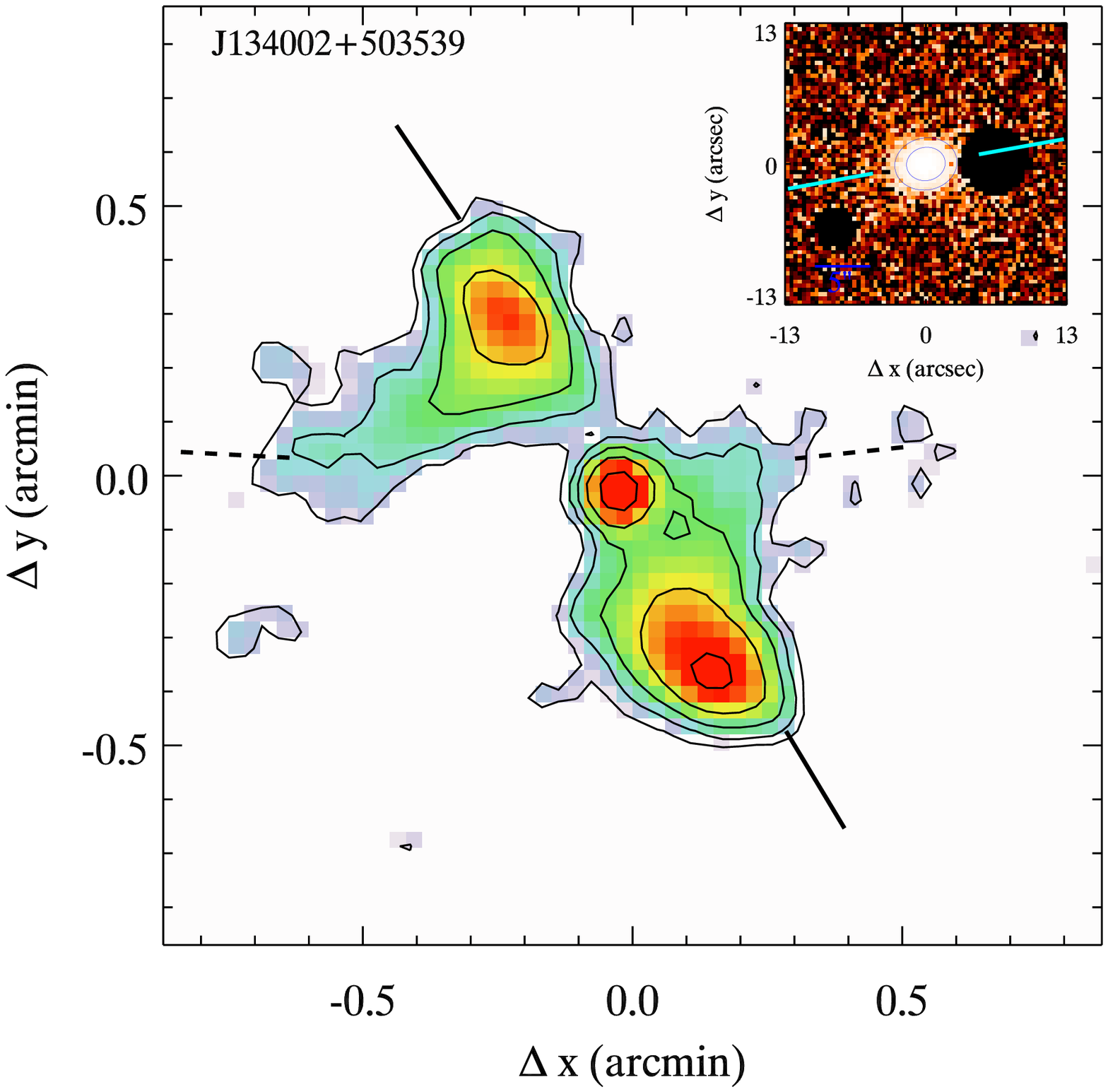} 
  \caption{The 6 X-shaped radio galaxies where the radio wings are found 
  to be aligned closer the optical major axis of the host elliptical. The
  inferred axes of the primary radio lobes and of the radio wings are
  marked with solid and dashed lines, respectively. The color images as 
  well as the contours refer to the 1.4 GHz FIRST maps (except
  for J000450$+$124839 for which the color image represents the 1.4GHz NVSS 
  map). The insets show the SDSS $i$-band image of the parent galaxies,
  whose optical major axes are marked with pairs of solid cyan lines.}
\label{fig:pa_back}
 \end{figure*}

\section{Optical versus Radio Structural Alignment}
\label{sec:opt_vs_radio_str}
To explore the geometric relationship between the radio structure and
the parent elliptical galaxy, the first step was to determine the
optical major axis of the elliptical. For this we mainly used the SDSS
$i$-band images. Among the 106 strong XRG candidates (taken here as
bona-fide XRGs), 7 sources have no optical object in the SDSS at the
location of the host galaxy expected from the symmetry consideration
of the radio lobes. For another 11 XRGs not covered in SDSS, we have
used the $r$-band images from the DECam Legacy Survey [DECaLS]
\citep{Dey2018arXiv180408657D}. Three XRGs are covered neither in SDSS
nor in DeCaLS. This left us with 96 XRGs with detected optical hosts.

For each of these XRGs, elliptical 
isophotes were fitted to the optical image using the {\sc iraf}
\footnote{IRAF is distributed by the National Optical Astronomy
  Observatory, which is operated by the Association of Universities
  for Research in Astronomy, Inc., under cooperative agreement with
  the National Science Foundation.} task {\tt`Ellipse'}
\citep{Jedrzejewski1987MNRAS.226..747J}. In brief, we fitted the
isophotes with sampling radius increasing exponentially in step of
1.1, and then characterized the galaxy ellipticity and position angle
as well as their respective uncertainties as the mean value and
standard deviation over the profiles in the outer part of the galaxy
image, where the intensity is about 1 $\sigma$ above the sky level. A
secure measurement of these quantities was often rendered difficult
owing to the rather small angular size of the ellipticals, which is
mainly due to their relatively large redshifts (median $z \sim 0.37$),
as  well as their rather shallow images in many cases. Therefore, we
imposed a condition that the ellipticity must be above 0.1 (at $> 2
\sigma$ level), which corresponds to the optical axial ratio $> 1.1$,
and, secondly, the estimated error on the measured position angle of
the optical main axis should be $\lesssim 10 $ degrees. Only 41 of the
XRGs could satisfy these conditions imposed in the interest of
reliability of the estimated parameters, as listed in columns 7 and 8
of Table~\ref{tab:radiopa_all} in Appendix A1. \par

Next, to estimate the position angle of the axes of the primary radio
lobes and the wings, we constructed polar diagrams of the FIRST radio
map at 1.4 GHz, taking the host galaxy as the origin (for details, see
\citet{Gillone2016A&A...587A..25G}. Using this diagram, we defined the
position angle of the two radio wings as well as the primary lobes as
the orientation of the wings (and the primary lobes) along the
directions having the greatest distance from the center at which the
brightness still exceeds the local 3$\sigma$ noise level. The two
examples shown in Fig.~\ref{fig:pa_img} (lower panels) display the
measured radio position angle (PA) for the wings and the primary
lobes, marked by dashed and solid lines, respectively. The upper
panels contain the polar diagram showing the radio contours of the
FIRST map in polar coordinates where the primary radio lobes produce
the two main peaks and the diffuse extended wings produce the well
defined linear features, indicating the position angle. The estimated
PAs of the axes of the primary and secondary lobe pairs are listed in
columns 3 and 5 of Table~\ref{tab:radiopa_all} in Appendix A1.

The afore-mentioned analysis has enabled examination of any
geometrical linkage between the properties of the host galaxy and the
associated X-shaped radio structure. Specifically, we have measured
the position angle offsets between the major axis of the optical host
(columns 3 and 4 of Table~\ref{tab:deltapa}) from the axis defined by
the two primary radio lobes (primary radio axis) and by the pair of
the radio wings (secondary radio axis). Figure~\ref{fig:pa} shows the
distributions of the two angular offsets. Note that we have taken
average of the axes of the two paired radio lobes. It is evident that
for the vast majority ($\sim 76\%$) of XRGs, the radio-optical
position angle offset for the secondary radio axis (defined by the two
wings) is larger than 30$^{\circ}$, with a median value of
57$^{\circ}$.  A Kolmogorov-Smirnov test (KS-test) rules out the
  null probability of being drawn from a uniform distribution at 
0.05  significance level. This trend, first reported by
\citet{Capetti2002A&A...394...39C} and most recently updated by
\citet{Gillone2016A&A...587A..25G}, are both based on relatively small
samples consisting of just 9 and 22 XRGs, respectively. The present
analysis of a larger sample has further strengthens the claims that
the radio wings in XRGs are preferentially aligned closer to the
optical minor axis of the host elliptical galaxy. Interestingly,
however, we do find 6 counter-examples, where the axis defined by the
radio wings lies closer to the optical {\it major} axis of the host
galaxy (see Fig.~\ref{fig:pa_back}). The existence of even a handful
of such counter-examples is interesting since it calls for caution in
accepting the hydrodynamical explanation (i.e., the backflow diversion
model) for the radio wings as the universal explanation for their
formation (Section~\ref{sec:intro_xrg}).

\begin{figure}
  \centering
\includegraphics[clip,height=8.2cm,width=8.2cm]{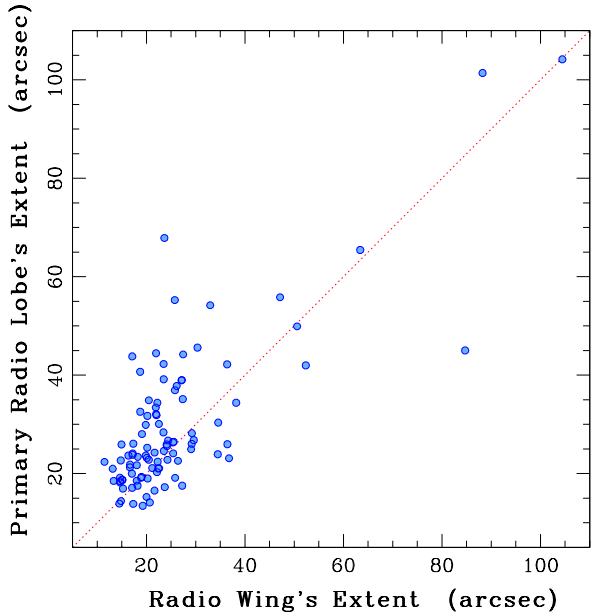}
  \caption{Distribution of the lengths of the primary and secondary lobes,
   for our sample of 106 XRGs. }
\label{fig:l_dist}
 \end{figure}

In Fig.~\ref{fig:l_dist}, we have plotted the radio extents of the
primary lobes and the associated wings for the 106 XRGs, as listed in
columns 4 and 6 of Table~\ref{tab:radiopa_all}. Note that the usual
definition of XRGs according to which the wing should extend at least
80\% of the size of the associated primary lobe had been relaxed while
constructing our XRG catalog \citep[see section
  2,][]{Yang2019arXiv190506356Y}. This was intended to make an
allowance for the fact that the detectability of the `tip' of the
radio wings can be hampered due to myriad factors, related to the
sensitivity limitation and also the source evolution, its orientation
and the wings' directional offset from the primary lobe axis. Here we
have plotted the larger of the two wings and of the two primary lobes
in Fig.~\ref{fig:l_dist}. It is seen that, taking an uncertainty of
$\sim 10\%$ in measuring radio size, the fraction of XRGs having wings
longer than the primary lobes can be as low as $\sim 20\%$ (the excess
factor is $>$ 1.5 in 3 cases). Note that in some of the XRGs, the
wings could appear longer than the primary lobes merely due to
 foreshortening of the latter caused by orientation to our line of sight. Accounting for this could
bring the apparent sizes more in tune with the basic backflow
diversion scenario wherein a wing is expected to be intrinsically
shorter than  the primary lobe. On the other hand, since we are
only concerned here with radio galaxies  and not quasars, any foreshortening of the
primary radio axis due to projection is not expected to be a major
effect (in statistical terms). At this stage, it appears that an XRG having
radio wings intrinsically longer than its active lobes is not a rare
occurrence. Clearly, a satisfactory resolution of this issue would
need deeper radio imaging at meter wavelengths.

 \begin{figure}
  \centering
\includegraphics[clip,height=8.2cm,width=8.2cm]{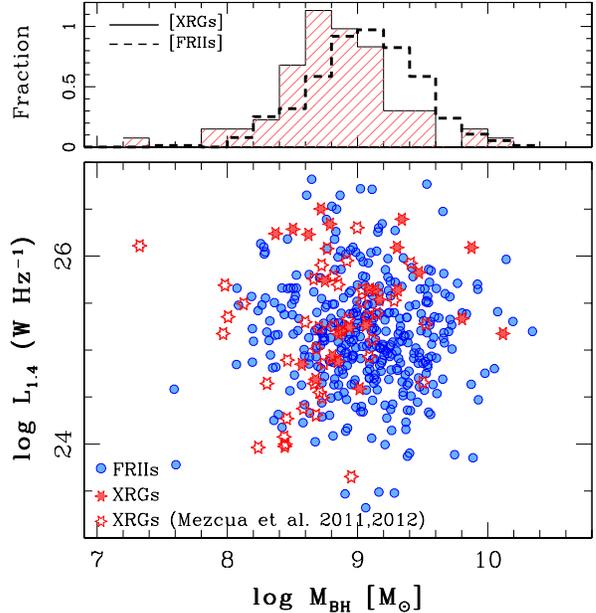}
  \caption{\emph{Lower panel:} Relation between black hole mass
  ($M_{\rm BH}$) and radio luminosity at 1.4 GHz for our 26 XRGs
  (filled stars), for the 41 XRGs from   \citet{Mezcua2011A&A...527A..38M,Mezcua2012A&A...544A..36M}
  (open stars) and for the 388 \frii radio galaxies (circles).
  \emph{Upper panel:} Distribution of $M_{\rm BH}$ found here for the 67 XRGs
  and 388 \frii radio galaxies.}  
\label{fig:mbh_dist}
 \end{figure}

 \begin{table}
 \centering
 \begin{minipage}{85mm}
 {\scriptsize
 \caption{Position angle offset of the axes defined by the pair of radio
 wings and by the two primary lobes, from the optical major axis of the 
 host elliptical galaxy, for the 41 XRGs in which it was possible to 
 measure all these parameters with reasonable accuracy.}
 \label{tab:deltapa}
 \begin{tabular}{@{} r c r  r  @{}}
 \hline  \hline 
  \multicolumn{1}{c}{Id}   &\multicolumn{1}{c}{Name}  &     \multicolumn{1}{c}{$\Delta$ PA$^{\textcolor{blue}{a}}$ }& \multicolumn{1}{c}{$\Delta$PA$^{\textcolor{blue}{b}}$} \\
  \multicolumn{1}{c}{  }   &\multicolumn{1}{c}{   }  &      \multicolumn{1}{c}{ passive 1/2 ($^{\circ}$)     } & \multicolumn{1}{c}{ active 1/2 ($^{\circ}$)  } \\
\hline 

  1   &J000450.27$+$124839.52 &       0/  11    &      47/   52   \\ 
  2   &J002828.94$-$002624.60 &      67/  86    &      76/   77   \\ 
  3   &J003023.86$+$112112.50 &       6/  25    &      35/   35   \\
  4   &J012101.23$+$005100.38 &      59/  55    &      30/   33   \\
  5   &J021635.79$+$024400.90 &      39/  26    &      74/   72   \\    
  6   &J031937.58$-$020248.70 &      44/  45    &       5/    7   \\ 
 11   &J075445.52$+$242425.30 &      15/  $-$   &      58/   69   \\ 
 15   &J081404.55$+$060238.38 &      72/  74    &       4/    6   \\ 
 16   &J081601.88$+$380415.48 &       5/  14    &      56/   68   \\ 
 17   &J081841.57$+$150833.50 &      80/  79    &      31/   13   \\ 
 19   &J082400.50$+$031749.30 &      84/  66    &      44/   59   \\ 
 21   &J084509.65$+$574035.54 &      68/  75    &      24/   15   \\ 
 27   &J092401.16$+$403457.29 &      24/  22    &      17/   24   \\ 
 28   &J092802.68$-$060752.63 &      64/  30    &      31/   29   \\ 
 29   &J093014.90$+$234359.20 &      70/  44    &      72/   71   \\ 
 32   &J094953.64$+$445655.77 &      99/  97    &      33/   33   \\ 
 37   &J103924.92$+$464811.53 &      52/  50    &      53/   62   \\ 
 38   &J104632.43$-$011338.15 &      57/  63    &      71/   60   \\
 40   &J112848.72$+$171104.57 &       1/  12    &      69/   75   \\ 
 42   &J114522.19$+$152943.26 &     110/  59    &      17/   10   \\ 
 44   &J115500.34$+$441702.22 &      92/  70    &       1/    1   \\ 
 45   &J120251.32$-$033625.80 &      46/  22    &      70/   69   \\ 
 50   &J131226.65$+$183414.98 &      13/   8    &      92/   82   \\ 
 51   &J131331.40$+$075802.51 &      62/  71    &      70/   78   \\
 52   &J132324.26$+$411515.01 &     114/  49    &      16/   13   \\ 
 53   &J132404.20$+$433407.14 &      54/  38    &     113/  112   \\ 
 54   &J132713.87$+$285318.19 &      45/  45    &      24/   14   \\ 
 56   &J133051.04$+$024843.10 &     117/  73    &      24/   23   \\ 
 58   &J134051.19$+$374911.74 &      45/  43    &     107/  106   \\ 
 59   &J134002.96$+$503539.72 &      11/   2    &      64/   67   \\ 
 69   &J150904.13$+$212415.10 &      62/  58    &      50/   50   \\ 
 71   &J151704.61$+$212242.14 &      49/   5    &      84/   84   \\ 
 76   &J155416.04$+$381132.57 &      99/  76    &      39/   44   \\ 
 77   &J160809.55$+$294514.92 &      53/  12    &      42/   43   \\ 
 78   &J162245.42$+$070714.69 &      42/  44    &       7/    7   \\ 
 79   &J164857.36$+$260441.26 &      52/  99    &      17/    4   \\ 
 81   &J202855.27$+$003512.67 &      29/  26    &      47/   44   \\ 
 86   &J223628.89$+$042751.89 &      72/  80    &      22/   31   \\ 
 87   &J232020.30$-$075319.36 &      92/  100   &      40/   43   \\ 
 99   &J125721.87$+$122820.58 &      60/  64    &       4/    9   \\ 
106   &J203459.54$+$005221.41 &      91/  86    &      29/   17   \\ 
                     
\hline               
 \end{tabular}       
 }  \\          
    $^{\textcolor{blue}{a}}$ Position angle difference between the optical major axis and passive lobes. \\
    $^{\textcolor{blue}{b}}$ Position angle difference between the optical major axis and active lobes. \\

 \end{minipage}
 \end{table}  
  
\section{Optical spectral properties}

We now examine certain physical parameters of XRGs, derived
from their optical spectra and then present a comparison with a large,
well-defined, sample of normal \frii radio galaxies. Our analysis is
based on the SDSS optical spectra which are available for 38 of our
106 XRGs. Six of these 38 had to be discarded as they are identified with 
quasars (including two showing double peaked narrow emission lines and 
no evidence for starlight). The reduced one-dimensional spectra for the 
remaining 32 XRGs and also for the control sample of 388 \frii radio 
galaxies were downloaded from the SDSS-BOSS Data Archive 
Server\footnote{http://dr12.sdss3.org/bulkSpectra}.

\subsection{Masses of the Central Black Holes} 
\label{subsec:bhmass}
To measure the required absorption and emission line parameters for
the 32 XRGs and the control sample of 388 \frii galaxies (see
Section~\ref{sec:sample}), we first corrected each spectrum for
reddening, taking the $E(B-V)$ values from
 \citet{Schlegel1998ApJ...500..525S}. This was followed up with
spectral fitting using the penalized PiXel-Fitting method (pPXF,
 \citealt{Cappellari2004PASP..116..138C,
  Cappellari2017MNRAS.466..798C}). Briefly, pPXF works in the pixel
space, and performs a nonlinear least-squares optimization to provide
the best-fit template and the velocity dispersion of the underlying
stellar population. The emission lines in the de-reddened spectrum
were masked and the underlying absorption spectrum was modeled as a
combination of single stellar population templates with {\sc miles}
 \citep{Vazdekis2010MNRAS.404.1639V} which are available for large
ranges in the metallicity [M/H] (from $\sim \rm -2.32\ to\ +0.22$) and
age (from $\rm 63\ Myr\ to\ 17\ Gyr$).

Among the 32 XRGs, six have average SNR per pixel of $<$ 3, which is
insufficient for a reliable measurement of  a stellar velocity dispersion.
Hence these could be obtained only for the remaining 26 XRGs, and are
listed in column 4 of Table~\ref{tab:spec}. We then proceeded with
the estimation of mass of the central SMBH, using the well known 
tight relation:
  \citep{Ferrarese2001ApJ...555L..79F,Gebhardt2000ApJ...539L..13G,
  Tremaine2002ApJ...574..740T}, given in
 \citet[][see their equation 3 and 7]{Kormendy2013ARA&A..51..511K}:

\begin{equation}
\rm  \frac{M_{\rm BH}}{10^9 M_{\odot}}  = \left(0.310^{+0.037}_{-0.033}\right) 
\left(\frac{\sigma_{\star}} {200\ {km\ s^{-1}}}\right)^{4.38\pm0.29}  
\label{eq1}
\end{equation}

In the upper panel of Figure~\ref{fig:mbh_dist} we compare the SMBH
mass distribution determined for the XRGs and for the control sample
of 388 \frii radio galaxies. In this comparison, we have also included
the 41 XRGs  known earlier for which stellar velocity dispersions are available in
 \citet{Mezcua2011A&A...527A..38M,Mezcua2012A&A...544A..36M}. The
average $M_{\rm BH}$ for the \frii radio sources is found to be
$M_{\rm BH} = 9.07 M_\odot$, which is slightly higher than the average
$M_{\rm BH} = 8.81 M_\odot$ estimated for the 67 XRGs \citep[see
  also,][]{Kuzmicz2017MNRAS.471.3806K}. A two-sided Kolmogorov-Smirnov
test shows that the population of XRGs and \frii radio sources are
drawn from different distributions with $KS-$test null probability of
$p_{null}$ = 0.005. In the lower panel of Fig.~\ref{fig:mbh_dist} we
look for any dependence of $M_{\rm BH}$ on the 1.4 GHz radio
luminosity, for our set of 67 XRGs. The Spearman rank-correlation test
does not show a significant correlation, with $r_s$ = 0.24 and a null 
probability of $p_{null}$ = 0.04.

 \begin{figure}
  \centering
\includegraphics[clip,height=8.2cm,width=8.2cm]{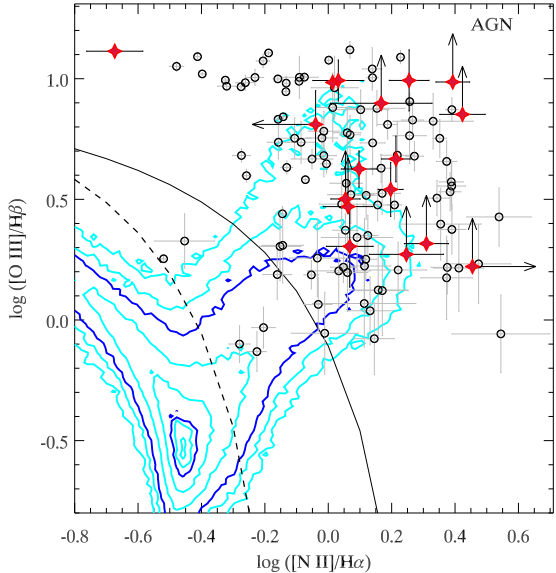}
  \caption{Diagnostic diagram: \o3hb versus \n2ha for XRGs (red stars)
    and \frii comparison sample (open circles). The solid black line
    indicates the boundary between AGNs and star-forming galaxies,
    following \citet{Stasinska2006MNRAS.371..972S}.}
\label{fig:bpt}
 \end{figure}

 \begin{figure}
  \centering
\includegraphics[clip,height=8.2cm,width=8.2cm]{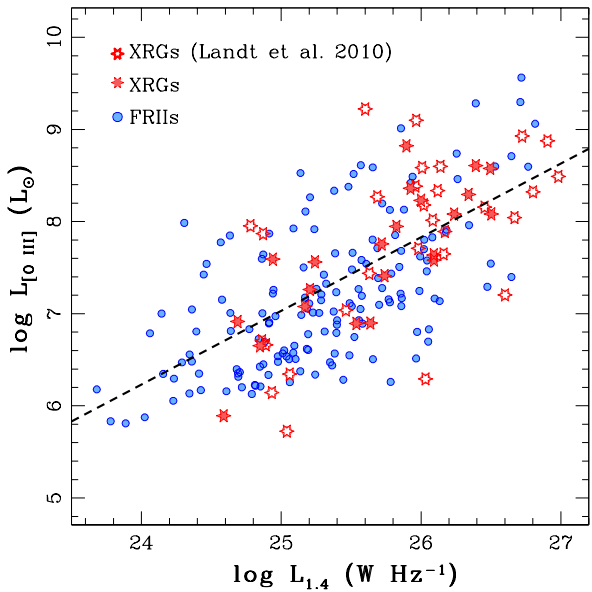}
  \caption{Radio luminosity at 1.4 GHz versus \oiii\ line luminosity
    ($\rm erg\ s^{-1}$) for our XRGs (filled stars), XRGs from
    \citet{Landt2010MNRAS.408.1103L} (stars) and the control sample of
    \frii radio galaxies (circles) by
    \citet{Koziel2011MNRAS.415.1013K}. The dashed line represents the
    best fit for the XRGs,  as discussed in the text.}
\label{fig:lvsp_dist}
 \end{figure}

\subsection{Radio and optical emission-line luminosities}
\label{subsec:xrgopt}
In order to measure the emission-line parameters we have modelled the
emission line profiles as Gaussians superposed on the continuum
subtracted spectrum. In brief, we first model the \sii\ doublet lines
which are taken to delineate the [N{\sc~ii}] and H$\alpha$ narrow
emission lines \citep{Greene2004ApJ...610..722G}. The satisfactory fit
to the line shape, thus obtained, was used as a template for other
narrow lines. Both the \sii\ doublet lines were assumed to have equal
widths (in velocity space) and be separated by their laboratory
wavelengths. A single Gaussian profile provided a good fit in cases of
all the XRGs, including the 16 XRGs where \sii\ lies outside the
spectral coverage of the SDSS. This analysis allowed us to place our
XRGs in the BPT diagnostic diagram
\citep{Kewley20062006MNRAS.372..961K} in which the intensity ratios of
\oiii/\hbeta are plotted against the ratios of [N{\sc~ii}]/\halpha
(Fig.~\ref{fig:bpt}). The value of these ratios, including upper
limits, could be measured only for 17 of the 32 XRGs. We also display
the same for a subset of 100 \frii radio galaxies (out of the control
sample of total 388) for which the emission lines have been detected
at $\ge 3 \sigma$ level. Interestingly, all the 17 XRGs and nearly all
of the 100 \frii galaxies are seen to inhabit the locus for AGNs
wherein the emission lines are excited predominantly due to AGN
activity. A similar trend has been noted by
\citet{Gillone2016A&A...587A..25G}, based on a smaller sample of 11
XRGs.

For 3C radio sources, \citet{Rawlings1989MNRAS.240..701R} have
reported a clear positive correlation between the luminosities of the
1.4 GHz emission ($L_{1.4}$) and the \oiii\ emission line (\loiii).
This points to the existence of a coupling between the physical
processes that power the narrow emission-line activity and the
extended radio synchrotron emission. This correlation has been confirmed
by \citet{Koziel2011MNRAS.415.1013K} for their larger set of 401
\frii radio galaxies. In Fig.~\ref{fig:lvsp_dist} we examine this 
for our XRG sample for which the values of \loiii\ and 1.4 GHz 
luminosities are listed in columns 7 and 9 of Table~\ref{tab:spec}. 
In the present analysis we have also included the 31 XRGs appearing 
in the catalog of 100 XRGs \citep{Cheung2007AJ....133.2097C}, for 
which \oiii\ line measurements are available in \citet{Landt2010MNRAS.408.1103L}. 
From Fig.~\ref{fig:lvsp_dist}, it is evident that a strong correlation
between \oiii\ line luminosity and 1.4 GHz radio luminosity exits for 
XRGs, as well, and we can parameterise it as: log L$_{[\rm OIII]}$ = 
[$0.80 \pm 0.14] $log $L_{1.4} - [12.97 \pm 3.58]$. The Spearman rank 
correlation test confirms the correlation with $r_s$ = 0.60 with a null 
probability, $p_{null} = 2.1 \times10^{-6}$ (i.e., significant at 4.3$\sigma$ level).

 Next,  we classify the radio galaxies into low- and high-excitation
 classes (LERGs and HERGs)  by following the criteria of \citet{Best2012MNRAS.421.1569B} 
 where a host galaxy with \oiiib line of equivalent-width $> 5$\AA\ is 
 classified as HERG. The \oiiib line equivalent-width and the corresponding 
 spectral classification are listed for our XRG sample in column 8 and 10 
 of Table~\ref{tab:spec}. Out of the 32 XRGs, 16 (i.e., 50 \%) are found 
 to be HERGs, and the remaining 16 XRGs, including the 9 XRGs with only 
 3$\sigma$ upper limit available for the \oiiib line equivalent-width, 
 fall in the category of LERGs.

 \begin{table*}
 \centering
 \begin{minipage}{180mm}
 {\scriptsize
 \caption{Optical spectral parameters of the X-shaped radio galaxies.} 
 \label{tab:spec}
 \begin{tabular}{@{} r c r r r r r c  c @{}}
 \hline  \hline 
  \multicolumn{1}{c}{Id}   &\multicolumn{1}{c}{Name}  &       \multicolumn{1}{c}{z} & \multicolumn{1}{c}{$\sigma_{\star}$} & \multicolumn{1}{c}{log$M_{\rm BH}$} &\multicolumn{1}{c}{log$L_{\rm [OIII]}$ }& \multicolumn{1}{c}{$EW_{\rm [OIII]}$} & \multicolumn{1}{c}{log L$_{1.4 GHz}$} & \multicolumn{1}{c}{spectral} \\
  \multicolumn{1}{c}{  }   &\multicolumn{1}{c}{   }  &    \multicolumn{1}{c}{ } &   \multicolumn{1}{c}{[\kms] }&   \multicolumn{1}{c}{[$M_{\odot}$] } & \multicolumn{1}{c}{ [erg s$^{-1}$]    } &\multicolumn{1}{c}{[\AA]  }& \multicolumn{1}{c}{$W Hz^{-1}$} & \multicolumn{1}{c}{ type}  \\
 \multicolumn{1}{c}{(1)}  &\multicolumn{1}{c}{ (2) }  &    \multicolumn{1}{c}{(3)} &   \multicolumn{1}{c}{(5)}&   \multicolumn{1}{c}{(6)} & \multicolumn{1}{c}{ (7)  } &\multicolumn{1}{c}{(8)}& \multicolumn{1}{c}{(9)} & \multicolumn{1}{c}{ (10)}  \\
\hline 

3 & J003023.85+112112.4 &    0.449&     $-$	    &        $-$      	   &    42.41$\pm$      0.01 &   153.36 $\pm$     2.25&24.69&  HERG\\
4 & J012101.23+005100.4 &    0.238& 281 $\pm$  46&       9.1 $\pm$     0.4 &    40.48$\pm$      0.07 &     2.84 $\pm$     0.86&25.21&  LERG\\
10& J072737.49+395655.5 &    0.312& 470 $\pm$  37&      10.1 $\pm$     0.3 & $<$40.34                &          $-$           &25.64&  LERG\\
17& J081601.88+380415.3 &    0.173& 286 $\pm$  18&       9.2 $\pm$     0.2 &    40.48$\pm$      0.11 &     2.09 $\pm$     0.84&25.54&  LERG\\
20& J082400.50+031749.3 &    0.215& 253 $\pm$  31&       8.9 $\pm$     0.3 &    41.15$\pm$      0.02 &     8.12 $\pm$     0.49&25.26&  HERG\\
21& J084509.65+574035.7 &    0.237& 241 $\pm$  25&       8.9 $\pm$     0.3 & $<$40.03                &  $<$3.74               &25.34&  LERG\\
22& J085236.12+262013.6 &    0.477&     $-$	    &        $-$           &    41.34$\pm$      0.03 &    15.39 $\pm$     1.27&25.24&  HERG\\
23& J085915.19+080539.7 &    0.565& 307 $\pm$  90&       9.3 $\pm$     0.6 & $<$40.69                &  $<$5.88               &26.41&  HERG\\
31& J092401.17+403457.1 &    0.160& 399 $\pm$  13&       9.8 $\pm$     0.2 & $<$40.18                &  $<$0.60               &24.94&  LERG\\
33& J093014.90+234359.1 &    0.538& 225 $\pm$  51&       8.7 $\pm$     0.5 &    41.67$\pm$      0.05 &    10.36 $\pm$     1.70&24.94&  HERG\\
34& J093238.29+161157.2 &    0.191& 249 $\pm$  23&       8.9 $\pm$     0.2 &    41.82$\pm$      0.01 &    39.93 $\pm$     0.59&26.00&  HERG\\
37& J095640.76$-$000124.0 &  0.139& 221 $\pm$  25&       8.7 $\pm$     0.2 &    40.50$\pm$      0.05 &     2.91 $\pm$     0.57&24.85&  LERG\\
38& J100408.95+350623.6 &    0.611& 312 $\pm$  81&       9.3 $\pm$     0.5 &    42.19$\pm$      0.01 &    41.59 $\pm$     1.60&25.74&  HERG\\
39& J101028.09+530313.2 &    0.341& 270 $\pm$  47&       9.1 $\pm$     0.4 & $<$40.46                &  $<$1.61               &25.17&  LERG\\
41& J101732.51+632953.7 &    0.184& 245 $\pm$  15&       8.9 $\pm$     0.2 &    40.85$\pm$      0.03 &     4.55 $\pm$     0.48&24.59&  LERG\\
44& J103900.86+354050.1 &    0.569&     $-$	    &        $-$           & $<$41.03                &  $<$5.35               &24.88&  HERG\\
45& J103924.92+464811.6 &    0.531& 414 $\pm$  35&       9.9 $\pm$     0.3 &    41.22$\pm$      0.06 &     4.62 $\pm$     0.92&25.65&  LERG\\
49& J112848.71+171104.6 &    0.347& 214 $\pm$  28&       8.6 $\pm$     0.3 & $<$40.26                &  $<$1.38               &26.23&  LERG\\
51& J113816.61+495025.3 &    0.510& 335 $\pm$  50&       9.5 $\pm$     0.4 &    41.53$\pm$      0.02 &    12.55 $\pm$     0.95&26.29&  HERG\\
52& J114522.18+152943.2 &    0.068& 263 $\pm$   7&       9.0 $\pm$     0.2 &    39.48$\pm$      0.11 &     0.63 $\pm$     0.29&26.09&  LERG\\
56& J122550.50+163343.4 &    0.656&     $-$	    &        $-$           &    42.16$\pm$      0.02 &    55.92 $\pm$     3.05&25.34&  HERG\\
57& J125721.88+122820.6 &    0.208& 229 $\pm$  37&       8.8 $\pm$     0.3 &    41.00$\pm$      0.02 &     6.80 $\pm$     0.48&25.93&  HERG\\
61& J130854.24+225822.1 &    0.677& 188 $\pm$  65&       8.4 $\pm$     0.6 & $<41.00$                &  $<4.9$                &25.93&  LERG\\
67& J132939.94+181841.9 &    0.514&     $-$	    &        $-$           &    41.67$\pm$      0.02 &    28.37 $\pm$     2.97&26.76&  HERG\\
69& J133051.02+024843.2 &    0.623& 201 $\pm$  64&       8.5 $\pm$     0.6 & $<$40.91                &  $<$3.59               &25.72&  LERG\\
70& J134002.96+503539.8 &    0.232& 242 $\pm$  17&       8.9 $\pm$     0.2 &    40.66$\pm$      0.06 &     3.79 $\pm$     1.01&24.80&  LERG\\
74& J140349.80+495305.4 &    0.491&     $-$	    &        $-$           &    41.95$\pm$      0.01 &    56.58 $\pm$     1.62&26.34&  HERG\\
84& J150904.13+212415.1 &    0.311& 275 $\pm$  27&       9.1 $\pm$     0.3 &    $<40.57$             &  $<$2.52               &25.64&  LERG\\
93& J160809.56+294514.8 &    0.226& 236 $\pm$  21&       8.8 $\pm$     0.2 &    41.18$\pm$      0.02 &    10.92 $\pm$     0.84&26.50&  HERG\\
95& J162245.42+070714.7 &    0.597& 234 $\pm$  73&       8.8 $\pm$     0.6 &    41.88$\pm$      0.03 &    22.62 $\pm$     1.79&26.50&  HERG\\
96& J164857.36+260441.1 &    0.137& 209 $\pm$  22&       8.6 $\pm$     0.3 &    40.24$\pm$      0.09 &     3.25 $\pm$     1.16&26.24&  LERG\\
104&J223628.90+042751.7 &    0.303& 306 $\pm$  29&       9.3 $\pm$     0.3 &    41.17$\pm$      0.02 &    12.33 $\pm$     0.92&26.23&  HERG\\                 
\hline               
 \end{tabular}       
}
 \end{minipage}
 \end{table*}

\begin{figure}
\centering
\includegraphics[clip,height=9cm,width=9cm]{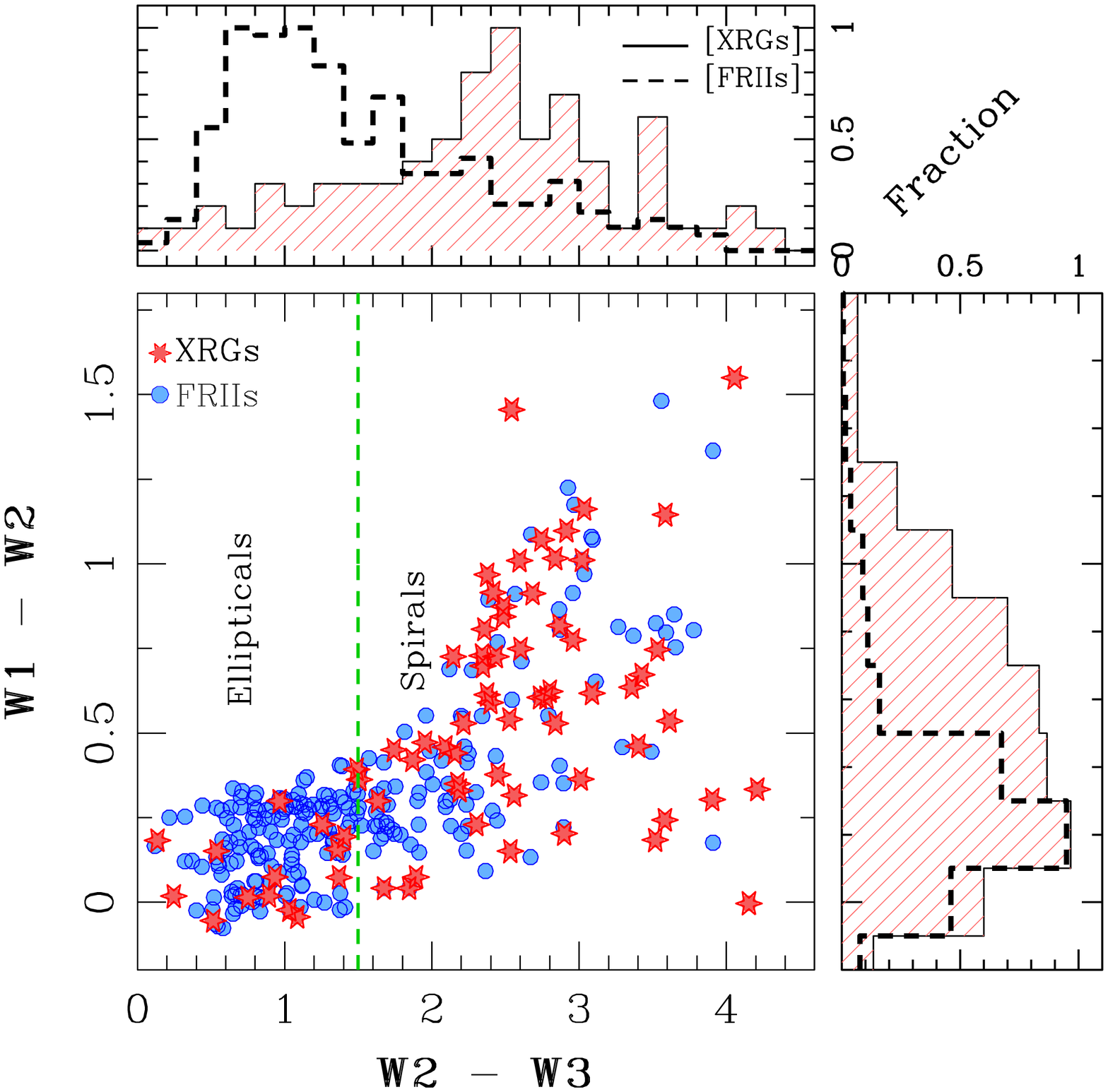}
\caption{The {\it WISE} color-color diagram drawn for XRGs (stars) and FRII radio 
galaxies (circles). The green dashed line with $W2-W3$ = 1.5 marks the 
division between elliptical and spiral galaxies
 \citep{Wright2010AJ....140.1868W}. The top and right-hand panels display
the distributions of $W2-W3$ and $W1-W2$ colors, respectively.}
\label{fig:xrg_wise}
 \end{figure}

\subsection{Infrared properties of XRGs}

We next investigate the mid-infrared properties of XRGs based on the
Wide-field Infrared Survey Explorer ({\it WISE}) archival information. For this, 
we have searched for the {\it WISE} counterparts to the host galaxies of 225 XRGs, 
including our  new 106 XRGs and the 119 XRGs assembled by \citet{Cheung2007AJ....133.2097C}. 
{\it WISE} counterparts could also be found for the entire control sample of 388 
\frii radio galaxies (note that only an offset of up to 3 arcsec between 
the SDSS and {\it WISE} positions was accepted). For each host galaxy, infrared 
colors were then calculated using their magnitudes given for the three mid-IR 
bands of the {\it WISE} survey [$W1 \rm (3.4\mu m), W2(4.6\mu m) and W3 (12\mu m)$
 in Vega magnitudes]. Fig.~\ref{fig:xrg_wise} shows the
color-color diagram for the 74 XRGs and 235 \frii radio galaxies,  which were
detected in all  three  {\it WISE} bands ($W1, W2$ and $W3$). 
It is noteworthy that the XRGs are typically found to be infra-redder than the
FRIIs, with a large fraction $\sim 80\%$ (59/74) actually falling in the 
region populated by spiral galaxies, i.e., the {\it WISE} $W2-W3$ color index
being $> 1.5 \rm~ mag$. In contrast, only $\sim 20\%$ of the \frii radio 
galaxies are found to inhabit that space. This indicates that compared to 
\frii radio galaxies, XRGs have a greater abundance of cool ISM, probably 
contributed by a recent merger. Note also that \citet{Stern2012ApJ...753...30S} 
have suggested that $W1-W2 \ge 0.8$ cut can be used to identify the most 
powerful AGNs with a high degree of reliability. Employing this clue, we 
find that 16 out of the 74 XRGs (i.e., $\sim 22\%$) plotted in
Fig.~\ref{fig:xrg_wise} host a powerful AGN. This fraction would 
rise to $40\%$ if a less stringent color cut of $W1-W2 \ge 0.6$ is adopted,
following \citet{Wright2010AJ....140.1868W}.\par

The {\it WISE} color data have also been found useful for classifying radio 
galaxies into low- and high-excitation classes (LERGs and HERGs)
 \citep[see,][]{Gurkan2014MNRAS.438.1149G}.  
 \citet{Sadler2014MNRAS.438..796S} have shown that nearly all the HERGs 
in their sample have the color index ($W2-W3 \gtrsim$ 2).
Using the 2712 nearby radio-luminous galaxies, \citet{Yang2015MNRAS.449.3191Y} 
have shown that LERGs and HERGs have different mid-infrared properties, 
such that in the {\it WISE} color-color diagram the two excitation classes are 
separated by 
color index $W1-W2 = 0.4$, albeit with a significant overlap between 
the two populations \citep[see, also][]{Pace2016ApJ...818...65P}. 
We have applied the above two selection filters, namely $W1-W2 < 0.4$ 
and $W2-W3 < 2$  to estimate the fraction 
of LERGs and found that 20 out of 74 ($\sim 30\%$) of the {\it WISE} detected 
XRGs are consistent with the LERG classification.  However, bearing in 
mind the fairly large uncertainty in this estimate \citep{Pace2016ApJ...818...65P}, 
the rather modest fraction of LERGs inferred here may not be inconsistent 
with the higher LERG fraction ($\sim 50\%$) inferred above for XRGs, from 
optical spectroscopic classification (see Section~\ref{subsec:xrgopt}),
and also by  \citet{Gillone2016A&A...587A..25G}, who found similar  fractions of LERGs and HERGs
among XRGs.

\subsection{The environmental influence}

 The origin of the \fri\ and \frii dichotomy among radio galaxies has been 
 debated  for a long time. According to one viewpoint, its origin is intrinsic, 
 related to the central SMBH and/or the physical conditions in the 
 accretion flow, such that FRIs are preferentially LERGs and FRIIs are 
 mostly HERGs
 \citep{Marchesini2004MNRAS.351..733M,Hardcastle2006MNRAS.370.1893H,Kauffmann2008MNRAS.384..953K}.
 In the alternative proposition, their difference is attributed to the 
 interactions of the jets with the gaseous environment of the host galaxy
  \citep{Kawakatu2009ApJ...697L.173K,Lin2010ApJ...723.1119L,Capetti2017A&A...601A..81C}.
 In this scenario, \fri\ radio galaxies are deemed to be powered by weaker 
 central engines and their (weaker) jets are influenced by the ambient 
 circum-galactic medium to a greater degree, such that their (less 
 powerful) jets propagating through a denser medium experience a greater 
 resistance, which then leads to an \fri\ morphology
 \citep{Gopal1988Natur.333...49G,Gopal2000A&A...363..507G,
 Gopal2001A&A...373..100G,Kaiser2007MNRAS.381.1548K}.
Several studies have confirmed a higher abundance of \fri\ radio 
galaxies in denser environment, as compared to \frii radio galaxies 
 \citep{Gendre2013MNRAS.430.3086G, Rodman2019MNRAS.482.5625R}.

We have measured the clustering environment around the XRGs and also
for the control sample of FRII radio galaxies, within a (projected)
radius of 1 Mpc and a (photometric) redshift interval of $\pm$ 0.04(1
+ $z$) around the spectroscopic (or, alternatively, photometric)
redshift, $z$, of a given radio galaxy. Within these spatial and
redshift bounds, we counted the number of SDSS galaxies with absolute
magnitudes brighter than $M_r = - 19 $ and determined the clustering
richness (N${^{- 19}_{1}}$) of the environment by subtracting the
number of background galaxies (brighter than $M_r = $-$ 19 $ found
within an annulus with inner and outer radii of 2.7 and 3.0  Mpc
measured from the radio source position
\citep{Wen2012ApJS..199...34W,Gendre2013MNRAS.430.3086G}. In order to
minimize the incompleteness in the galaxy counts arising from the
adopted magnitude limit, as dictated by the SDSS photometric data, as
well as the steeply growing uncertainty in redshift estimation at
higher redshifts, we only selected galaxies having photo-$z < 0.4$.
This search, based on the SDSS archival data, covered all the 225
XRGs, including our 106 `strong' cases and the 119 XRGs assembled by
\citet{Cheung2007AJ....133.2097C}, as well as our control sample of
388 FRII radio galaxies (see above). We found 107 XRGs and 343 FRII
radio galaxies that satisfied the above selection filters of $z < 0.4$
and $M_r \le -19$, and determined the clustering environment around
them, following the aforesaid procedure. \par

\begin{figure}
\includegraphics[clip,height=8.5cm,width=8.5cm]{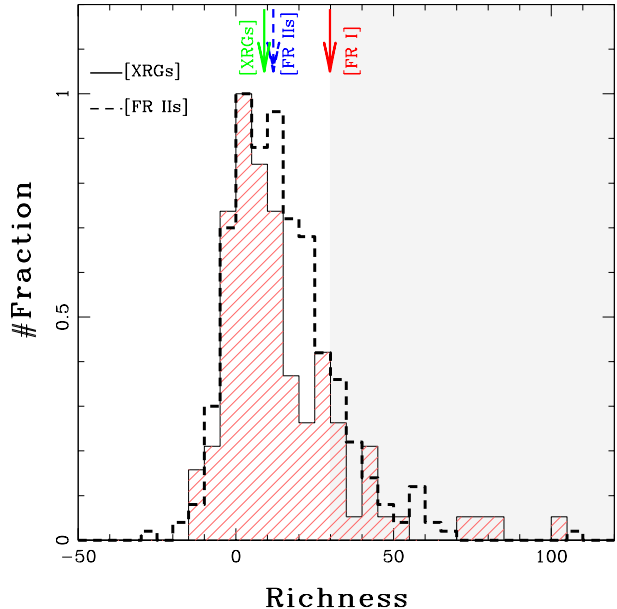}
  \caption{ Distribution of the environmental richness factor for XRGs 
  (hatched histogram) and \frii radio galaxies (dashed curve). The median
  richnesses for the XRGs and \frii are marked as solid and dashed arrows, 
  respectively, together with the median estimated richness for \fri\ radio 
  galaxies \citep{Gendre2013MNRAS.430.3086G}. The gray shaded region 
  corresponds to rich clustering environment, with richness $> 30$, the 
  boundary  taken to separate poorer from richer clusters \citep{Gendre2013MNRAS.430.3086G}.}
\label{fig:xrg_env}
 \end{figure}

 The derived distributions of the cluster richness for the sample of
 XRGs and the control sample of FRII radio galaxies are compared in
 Fig.~\ref{fig:xrg_env}. It is evident that XRGs and FRIIs inhabit
 similar weak clustering environments, the median cluster
 richnesses being 8.94 and 11.87, respectively. Both values correspond
 to a `poor' cluster environment. The shaded gray region in
 Fig.~\ref{fig:xrg_env} marks the N${^{- 19}_{1}} > 30$, normally
 regarded as the rough boundary separating poor from rich clusters
 \citep{Gendre2013MNRAS.430.3086G}. The present analysis
 shows that, very similarly, the vast majority of XRGs (91 out of 107,
 i.e., $\sim$85\%) and of the \frii radio galaxies (291 out of 343,
 i.e., $\sim$84\%) reside in poor environments. A similar median
 richness of 14.9  has been found by \citet{Gendre2013MNRAS.430.3086G} 
 for \frii radio sources. They also have confirmed that \fri\ sources 
 reside in relatively dense environments, with median richness of 29.8, 
 seemingly consistent with the premise that the (weaker) jets of \fri\ 
 sources, propagating in denser media, are more prone to disruption
 (see above). Thus, the preponderance of XRGs in poor clustering 
 environment appears in tune with the finding that they mostly belong 
 to the \frii class \citep[see section 5 of][]{Saripalli2009ApJ...695..156S}, and also
 \citet{Kuzmicz2019A&A...624A..91K}.

\section{Discussion and Conclusion}
\label{sec:dis}
In this study we have investigated the nature of X-shaped radio
galaxies (XRGs) by examining possible connections between their radio
and optical morphologies, their optical spectral properties as well as the
clustering environment around their host galaxies. The principal
trigger for this study came from a catalog of XRGs, recently published
by some of us \citep{Yang2019arXiv190506356Y} which contains 106
`strong' XRG candidates. We have used it for examining the geometrical
relationship between the axes of their radio structure and the
apparent orientation of the parent elliptical galaxies. This was
possible for 41 of the XRGs for which the position angle of the
optical major axis of the parent elliptical galaxy could be determined 
with a fair degree of reliability, primarily using their SDSS $i-$band 
images. 

The position angles of the primary radio axis (defined by the two
active lobes) and the secondary radio axis (defined by the two wings)
were measured by analyzing their FIRST 1.4 GHz maps, aided by the TGSS
150 MHz maps (Section~\ref{sec:opt_vs_radio_str}). This analysis has 
confirmed the previously reported strong tendency for the secondary 
radio axis to be closer to the optical minor axis (see the lower panel of 
Fig.~\ref{fig:pa}). Quantitatively, the directional offset between the
wings' axis from the optical major axis is found to exceed 30$^{\circ}$
(median 56$^{\circ}$) in $\sim 68\%$ of the XRGs.  The $KS-$test rejects the null
probability of being drawn from a uniform distribution at 
  0.05 significance level. Thus, our analysis strengthens
the earlier findings based on smaller samples of XRGs (e.g., by
\citealt{Capetti2002A&A...394...39C} and
\citealt{Gillone2016A&A...587A..25G}), which is basically in accord with 
the scenario that the radio wings in XRGs probably arise from diversion
of the back-flowing synchrotron plasma of the two primary lobes, by an
asymmetric circum-galactic medium of the host galaxy
(Section~\ref{sec:intro_xrg}). This is also seen in the recent 3D 
relativistic magnetohydrodynamic simulations of the twin-jets propagating 
in a triaxial density distribution, by \citet[][]{Rossi2017A&A...606A..57R}, 
see also, \citet[][]{Hodges2011ApJ...733...58H}. We note, however, this 
does not preclude alternative scenarios proposed for the origin of the 
wings, e.g., the spin-flip mechanism (Section~\ref{sec:intro_xrg}). This 
possibility is also favored by the finding that a non-negligible fraction 
(6 out of 41) of XRGs are counter-examples to this trend, wherein the 
secondary radio axis, defined by the wings, lies closer to the optical 
main axis of the parent elliptical galaxy (Section~\ref{sec:opt_vs_radio_str}). 
Earlier, only one such counter-example had been reported by \citet{Gillone2016A&A...587A..25G} 
(1 out of 22 XRGs in their sample). However, in that source even the sole 
visible radio wing has a very low surface brightness, rendering its 
position angle estimate quite uncertain. For the present, at least for 
the counter-examples, an alternative mechanisms (Section~\ref{sec:intro_xrg}), 
like the spin-flip scenario (or some variant of it) could offer a greater 
promise.

In the jet-shell interaction scenario, a ({\it transverse}) radio wing
develops as the jet's head gets nearly stalled temporarily due to its
interception by a (rotating) shell segement. The synchrotron plasma
deposited at the jet's head would then undergoes a (transverse)
diversion due to dragging by the shell, until the jet's blocking is
over and its rapid advance resumes (GBGW12). A plausible example of
this is witnessed in the radio galaxy 3C433 \citep[][see Fig. 1 of
  GBGW12]{Miller2009ApJ...695..755M}. Since a similar interruption is
likely to be experienced by the counter-jet (albeit not necessarily
contemporaneous to the main jet), a Z-symmetric morphology of the
radio wings can be expected to result. Note that in this jet-shell
interaction scenario, the wing's extent is mainly determined by the
duration for which the jet remains quasi-stalled by the shell, as
compared to the subsequent phase during which the jet propagates
freely ahead. It is then easily conceivable that the wing becomes
longer than the primary lobe, as indeed found in roughly a fifth of
XRGs (Fig.~\ref{fig:l_dist}) to the detriment of the basic backflow
diversion scenario. 

Another important ramification of this scenario is
that the launching point of the wing (due to transverse diversion of
the jet's synchrotron plasma) would coincide with the arresting of
the jet's head by the intercepting shell. Since this could well take
place dozens of kiloparsec away from the host galaxy, the existence of
large radio wings at such huge radial distances from the parent galaxy
can be readily understood, unlike the model which attributes the wing
formation to diversion of the backflowing plasma in the primary lobes,
by a non-spherical interstellar medium of the parent galaxy
(Section~\ref{sec:intro_xrg}). The existence of several cases of wings
originating so far away from the galaxies, has recently prompted the
classification of the wings into two types
\citep{Saripalli2018ApJ...852...48S}. According to these authors,
wings are launched from the active lobes at two `strategic' locations,
such that the preferred launching points are either fairly close to
the parent galaxy (i.e., within its ISM), or, alternatively just
somewhat behind jet's head (i.e., hot spot). They have termed the two
types as the inner (`I-dev') and outer (`O-dev') type wings. Whereas
the I-dev type wings, quite likely involve interaction with the ISM of
the host galaxy, which therefore defines their physical scale, the
preferred location of the `O-dev' type wings seems enigmatic. However,
this may be easily understood in the jet-shell interaction model for
the wing formation.
In this scenario, the growth of the wing peaks during the phase when the 
growth of the primary lobe remains muted as the jet's head remains nearly 
stalled by the intercepting shell sedgement (and, consequently, the backflow
of the synchrotron plasma intensifies). We propose that the occurrence of 
the two processes in tandem could explain the propensity of the `O-dev' 
type wings  to be found to be footed only a short way behind the jet head (see  above).

The `backflow diversion' scenario may also hold a clue to the intriguing result 
that in several XRGs, radio spectral variations are too small to reveal any 
sign of radiative aging in the wings  \citep[see,][and refs. therein]{Lal2019arXiv190311632L}. 
As the younger synchrotron plasma backflowing in the primary lobes, continues 
to be diverted into the existing cavities filled by the (low density) 
synchrotron plasma of the wings, this would keep the wings supplied with 
younger synchrotron plasma, thereby prolonging their radiative lifespans 
(see, also GBGW12; \citealt{Hodges2012ApJ...746..167H,Saripalli2018ApJ...852...48S}).

Comparing the mid-infrared properties of the 74 XRGs and a control
sample of 235 FRII radio galaxies with their counterparts in the {\it WISE}
catalog, we have shown that a large fraction $\sim 80\%$ (59/74) of
XRGs have redder hosts, with $W2-W3 > 1.5$. In contrast, only $\sim 20$\% 
of the FRII radio galaxies are found to inhabit that colour space 
(Fig.~\ref{fig:xrg_wise}). This mid-IR colour difference is indicative of
a higher abundance of dusty ISM in XRGs, possibly contributed by a recent 
galaxy merger. A similar effect has been noticed for double-double radio 
galaxies \citet[][]{Kuzmicz2017MNRAS.471.3806K}. Thus, galaxy merger appears 
to be  an especially important aspect of both these classes of radio galaxies 
and should feature in their theoretical modelling.\par

One arena where XRGs display striking similarity to normal \frii radio 
galaxies is the large-scale clustering environment around their host 
galaxies. In this work, we have compared the galaxy clustering densities 
around 107 XRGs and 343 FRII radio galaxies at $z < 0.4$, counting the 
$M_r \ge -19$ galaxies seen within 1 megaparsec and within a redshift 
interval of $\Delta z = 0.04(1 + z)$. Our analysis shows that both XRGs 
and FRIIs inhabit relatively poor environments, with median cluster richness 
of 8.9 and 11.87, respectively \citep[also see,][]{Gendre2013MNRAS.430.3086G}. 
In comparison, FRI radio galaxies with less well collimated radio jets 
are known to inhabit significantly richer environment, with median richness 
of $\sim 30$ \citep[see,][]{Gendre2013MNRAS.430.3086G}. This clearly 
reflects the role played by the clustering environment in the XRG phenomena.

In summary, the present study shows that the remarkable diversity in the 
observed properties of XRGs is hard to encompass within a single physical 
mechanism (i.e, backflow diversion via over-pressured cocoons, or a rapid 
reorientation of the AGN axis, or the jet-shell interaction scenario). It
appears that several mechanism are at work, and their relative importance 
may differ among the sources.  In particular, a clearer insight  would require
deeper optical and possibly X-ray imaging of the parent galaxies of XRGs, 
combined with their radio spectral mapping.

\section*{Acknowledgments}


We thank the anonymous referee for constructive comments
and suggestions.
This work was supported by the China Postdoctoral Science Foundation
Grants (2018M630024, 2019T120011), the National Key R\&D Program of
China (2016YFA0400702, 2016YFA0400703) and the National Science
Foundation of China (11473002, 11721303, 11533001). SS acknowledges
support from the Science and Engineering Research Board, India,
through the Ramanujan Fellowship.

Funding for the Sloan Digital Sky Survey IV has been provided by the
Alfred P. Sloan Foundation, the U.S. Department of Energy Office of
Science, and the Participating Institutions. SDSS-IV acknowledges
support and resources from the Center for High-Performance Computing
at the University of Utah. The SDSS web site is www.sdss.org.

SDSS-IV is managed by the Astrophysical Research Consortium for the
Participating Institutions of the SDSS Collaboration including the
Brazilian Participation Group, the Carnegie Institution for Science,
Carnegie Mellon University, the Chilean Participation Group, the
French Participation Group, Harvard-Smithsonian Center for
Astrophysics, Institutes de Astrof\'isica de Canarias, The Johns
Hopkins University, Kavli Institute for the Physics and Mathematics of
the Universe (IPMU) / University of Tokyo, Lawrence Berkeley National
Laboratory, Leibniz Institut f\"ur Astrophysik Potsdam (AIP),
Max-Planck-Institut f\"ur Astronomie (MPIA Heidelberg),
Max-Planck-Institut f\"ur Astrophysik (MPA Garching),
Max-Planck-Institut f\"ur Extraterrestrische Physik (MPE), National
Astronomical Observatories of China, New Mexico State University, New
York University, University of Notre Dame, Observat\'ario Nacional /
MCTI, The Ohio State University, Pennsylvania State University,
Shanghai Astronomical Observatory, United Kingdom Participation Group,
Universidad Nacional Aut\'onoma de M\'exico, University of Arizona,
University of Colorado Boulder, University of Oxford, University of
Portsmouth, University of Utah, University of Virginia, University of
Washington, University of Wisconsin, Vanderbilt University, and Yale
University. 

\bibliographystyle{aasjournal} 
\bibliography{ms}
\newpage

\section{Appendix}
\begin{center}
\renewcommand{\thefootnote}{\fnsymbol{footnote}}
{\scriptsize
\begin{longtable*}{ r r r  r r r  r   r c c}
\caption{Radio and  optical parameters of X-shaped radio galaxies.} \\
\hline 
\multicolumn{1}{c}{Id}   &
\multicolumn{1}{c}{Name} &
\multicolumn{2}{c}{Active-lobe}&
\multicolumn{2}{c}{Passive-lobe} &
\multicolumn{1}{c}{Optical} &
\multicolumn{1}{c}{Ellipticity} &
\multicolumn{1}{c}{flag$^{\textcolor{blue}{b}}$} \\
\multicolumn{1}{c}{  }   &
\multicolumn{1}{c}{   }  &
\multicolumn{1}{c}{ PA$^{\textcolor{blue}{a}}$($^{\circ}$)} &
\multicolumn{1}{c}{ length ($''$)}  &
\multicolumn{1}{c}{PA ($^{\circ}$)}  &
\multicolumn{1}{c}{  length ($''$)} &
\multicolumn{1}{c}{ PA ($^{\circ}$)} & 
\multicolumn{1}{c}{  }    &
\multicolumn{1}{c}{  }    \\
\hline
\endfirsthead
\multicolumn{9}{c}
{{\bfseries \tablename\ \thetable{} -- continued from previous page}} \\
\hline 
\endhead
\hline \multicolumn{9}{r}{{Continued on next page}} \\
\endfoot
\hline \hline
\endlastfoot
\label{tab:radiopa_all}
   1&J000450.27+124839.52&   286/    101&   172.4/ 148.9  &   53/  245&  126.1/  133.6 &  53  $\pm$   12 &     0.153  $\pm$   0.031  &  1  \\
   2&J002828.94$-$002624.60&  47/    226&    80.0/ 104.2  &   15/  214&   83.2/  104.4 & 123  $\pm$    9 &     0.199  $\pm$   0.038  &  1  \\
   3&J003023.86+112112.50&    61/    241&    39.2/  38.9  &   19/  180&   20.5/   23.5 &  25  $\pm$    7 &     0.271  $\pm$   0.016  &  1  \\
   4&J012101.23+005100.38&   360/    178&    23.4/  21.1  &   90/  266&   18.2/   12.0 &  30  $\pm$    7 &     0.397  $\pm$   0.071  &  1  \\
   5&J021635.79+024400.90&    12/    190&    62.1/  67.9  &   80/  272&   23.6/   18.9 & 118  $\pm$    9 &     0.188  $\pm$   0.035  &  1  \\
   6&J031937.58$-$020248.70& 291/    113&    17.5/  14.9  &  344/  163&   27.2/   24.2 & 118  $\pm$    5 &     0.301  $\pm$   0.032  &  1$^{\dagger}$  \\
   7&J071031.14+354649.80&     8/    185&    22.1/  25.3  &  316/  139&   18.3/   20.2 &  $-$                  &       $-$  	     &  0$^{\dagger}$  \\
   8&J071510.12+491053.28&   291/    123&    17.7/  18.6  &   26/  213&   13.4/   18.0 & 138  $\pm$    3 &     0.084 $\pm$    0.037  &  0  \\
   9&J072014.66+403748.68&   291/    110&    16.9/  18.9  &   30/  202&   13.1/   20.2 &  87  $\pm$    7 &     0.079 $\pm$    0.074  &  0  \\
  10&J072737.48+395655.84&   255/     45&    23.7/  11.1  &  306/  220&   19.9/   14.6 &         $-  $         &           $-  $     &  2  \\
  11&J075249.10+325254.20&    12/    180&    20.9/  24.6  &  288/   71&   23.5/   18.4 &  50  $\pm$    3 &     0.047 $\pm$    0.048  &  0  \\
  12&J075445.52+242425.30&    35/    204&    17.2/  12.0  &  305/  108&   14.4/   23.7 &  93  $\pm$    2 &     0.218  $\pm$   0.025  &  1  \\
  13&J075930.94+124722.86&     8/    185&    20.5/  21.7  &  298/  123&   10.1/   18.0 &  $-$                  &   $-$  	     &  3  \\
  14&J080006.84+495755.06&   296/    123&    16.8/  19.1  &   169/  188&   20.3/   25.8 &  $-$                  &   $-$  	     &  4  \\
  15&J081337.78+300710.60&   283/     90&    20.8/  21.8  &   21/  220&   16.7/   15.1 &  $-$                  &   $-$  	     &  3  \\
  16&J081404.55+060238.38&    29/    198&    23.6/  24.9  &  312/  130&   21.6/   29.1 &  24  $\pm$    5 &     0.143  $\pm$   0.050  &  1  \\
  17&J081601.88+380415.48&    71/    263&    22.8/  21.1  &    9/  180&   18.4/   20.5 &  14  $\pm$    2 &     0.372  $\pm$   0.011  &  1  \\
  18&J081841.57+150833.50&   314/    116&    14.4/  17.1  &   46/  225&   14.6/   17.1 & 146  $\pm$    2 &     0.194  $\pm$   0.042  &  1  \\
  19&J082226.42+051951.16&    60/    230&    31.9/  27.0  &  356/  135&   22.0/   21.9 &  $-$                  &   $-$  	     &  0  \\
  20&J082400.50+031749.30&    57/    252&    20.7/  23.9  &  307/  108&   34.5/   23.5 &  12  $\pm$    4 &     0.420  $\pm$   0.010  &  1  \\
  21&J084509.65+574035.54&     9/    180&    13.3/  13.1  &  277/   90&   15.0/   15.0 & 165  $\pm$    3 &     0.296  $\pm$   0.032  &  1  \\
  22&J085236.12+262013.41&   315/    135&    23.7/  22.7  &   515/  249&   13.1/   16.3 &  $-$                  &   $-$  	     &  0  \\
  23&J085915.19+080539.72&    84/    262&    14.9/  13.0  &    8/  110&   14.8/   18.1 &  $-$                  &   $-$  	     &  0  \\
  24&J085942.66+585116.64&    22/    213&    20.3/  18.4  &  290/   98&   22.1/   12.9 &  53  $\pm$    9 &     0.223 $\pm$    0.044  &  2  \\
  25&J085954.12$-$025241.93& 347/    161&    17.6/  22.6  &  287/  110&   26.4/   19.1 & 160  $\pm$    3 &     0.088 $\pm$    0.028  &  0$^{\dagger}$ \\
  26&J090331.03+560039.99&    80/    251&    11.4/  19.2  &  350/  178&   14.6/    8.2 &  $-$                  &   $-$  	     &  0  \\
  27&J090638.35+064524.62&   290/    180&    18.4/  15.8  &   90/  231&   14.7/   12.9 &  $-$                  &   $-$  	     &  0  \\
  28&J090827.86+215823.81&    34/    185&    21.0/  20.8  &  293/  135&   15.0/   22.4 &  $-$                  &   $-$  	     &  0  \\
  29&J091451.07+082440.17&   304/    128&    13.5/  12.9  &  60/  238&   12.9/   19.2 &  $-$                  &   $-$  	  	     &  0  \\
  30&J092346.43+361407.33&   290/     84&    19.3/  18.6  &  338/  $-$  &   19.0/    $-$ &  $-$                  &   $-$  	     &  0  \\
  31&J092401.16+403457.29&    45/    232&    23.1/  21.7  &    3/  185&   36.8/   24.4 &  27  $\pm$    3 &     0.319  $\pm$   0.025  &  1  \\
  32&J092802.68$-$060752.63& 275/     93&    30.4/  30.1  &    8/  150&   12.9/   34.6 & 124  $\pm$    2 &     0.178  $\pm$   0.019  &  1$^{\dagger}$ \\
  33&J093014.90+234359.20&    43/    223&    33.0/  43.8  &  339/  185&   17.1/   17.0 & 114  $\pm$    4 &     0.326  $\pm$   0.043  &  1  \\
  34&J093238.30+161157.33&   308/    128&   136.3/ 125.6  &  286/  101&   79.3/   70.7 &   7  $\pm$    3 &     0.054 $\pm$    0.064  &  2 \\
  35&J094240.45+044423.10&   290/    112&    18.5/  15.6  &   28/  231&    9.8/   13.4 &  $-$               &   $-$  	             &  0  \\
  36&J094953.64+445655.77&   328/    147&    54.2/  51.8  &   14/  196&   32.9/   29.0 & 113  $\pm$    9 &     0.165  $\pm$   0.017  &  1  \\
  37&J095640.77$-$000123.99& 323/    140&    63.3/  65.4  &    1/  200&   48.6/   63.4 &  54  $\pm$   38 &     0.017 $\pm$    0.023  &  0  \\
  38&J100408.95+350623.69&    52/    249&    31.7/  17.7  &    6/  157&   20.2/   14.8 &  $-$                  &   $-$  	     &  0  \\
  39&J101028.07+530313.06&    $-$           &     $-$         &    $-$    &    $-$   	        &   3  $\pm$  160 &     0.094 $\pm$  0.049  &  0  \\
  40&J101134.80$-$060753.14& 360/    206&    13.9/  13.4  &  325/  140&   17.3/   14.7 &  $-$                  &   $-$  	     &  0$^{\dagger}$ \\
  41&J101732.51+632953.82&   290/    119&    22.3/  24.1  &   62/  228&   23.5/   25.4 &         $-  $         &           $-  $     &  2  \\
  42&J103118.85+044307.70&    56/    226&    28.0/  26.8  &  2905/  127&   18.8/   19.1 &  $-$                  &   $-$  	     &  0  \\
  43&J103358.55+353007.24&    11/    170&    15.6/  16.9  &   80/  260&   11.7/   15.2 &  26  $\pm$    1 &     0.148 $\pm$    0.143  &  0  \\
  44&J103900.86+354050.12&    73/    255&    31.0/  34.9  &  288/  116&   20.5/   13.0 &         $-  $         &           $-  $     &  0  \\
  45&J103924.92+464811.53&   315/    126&    24.4/  25.9  &   60/  238&   36.4/   23.4 &   8  $\pm$    3 &     0.271  $\pm$   0.067  &  1  \\
  46&J104632.43$-$011338.15&  27/    219&    32.6/  29.5  &   90/  264&   16.6/   18.7 & 147  $\pm$    3 &     0.180  $\pm$   0.007  &  1  \\
  47&J105426.39+470327.47&   336/    158&    36.7/  38.9  &  286/  118&   27.1/   27.0 &   $-$                 &   $-$  	     &  0  \\
  48&J110853.80+263650.20&    45/    248&    18.6/  19.1  &  297/  110&   10.3/   18.9 &  $-$                  &   $-$  	     &  0  \\
  49&J112848.72+171104.57&    43/    217&    32.4/  44.4  &  279/  110&    5.5/   22.0 & 111  $\pm$    3 &     0.167  $\pm$   0.027  &  1  \\
  50&J113649.98+015121.34&    69/    252&    17.3/  24.3  &    5/  188&   21.6/   13.2 &  78  $\pm$   10 &     0.097 $\pm$    0.083  &  0  \\
  51&J113816.62+495025.03&   336/    140&    11.1/  14.1  &   70/  225&   20.7/   13.1 &  12  $\pm$    3 &     0.037 $\pm$    0.050  &  0  \\
  52&J114522.19+152943.26&    66/    239&    18.9/  22.8  &  288/  159&   24.3/   17.1 &  48  $\pm$    3 &     0.179  $\pm$   0.024  &  1  \\
  53&J115225.55+201602.19&    73/    249&    26.3/  17.2  &   48/  180&   25.4/   13.1 &  98  $\pm$   15 &     0.062 $\pm$    0.074  &  0  \\
  54&J115500.34+441702.22&    725/   252&    28.9/  36.9  &  324/  166&   25.8/   23.7 &  73  $\pm$    3 &     0.265  $\pm$   0.006  &  1  \\
  55&J120251.32$-$033625.80& 293/    114&    19.3/  21.1  &  358/  201&   21.2/   17.0 &  43  $\pm$    5 &     0.140  $\pm$   0.028  &  1  \\
  56&J122550.51+163343.50&   284/    1065&    39.0/  29.2  &  328/ 164&   18.5/   27.2 &         $-  $         &           $\pm$     &  0  \\
  57&J125721.87+122820.58&   333/    167&    45.0/  44.1  &   42/  226&   14.4/   84.7 & 162  $\pm$    9 &     0.202  $\pm$   0.033  &  1  \\
  58&J125900.79+203248.63&   355/    180&    20.9/  20.7  &   45/  244&   11.7/   13.2 &  $-$                  &   $-$  	     &  0  \\
  59&J130048.34+350527.35&   287/    119&    26.8/  24.2  &  310/  180&   29.6/   25.7 &  $-$                  &   $-$  	     &  3  \\
  60&J130258.46+511943.69&   293/    120&    42.1/  49.9  &   47/  201&   50.6/   25.4 &  $-$                  &   $-$  	     &  0  \\
  61&J130854.25+225822.30&    72/    242&    23.8/  23.7  &  306/  130&   17.2/   14.9 &  $-$                  &   $-$  	     &  0  \\
  62&J131226.65+183414.98&    55/    225&    17.0/  18.2  &  315/  157&    6.2/   14.6 & 143  $\pm$    3 &     0.557  $\pm$   0.016  &  1  \\
  63&J131331.40+075802.51&   341/    1545&    27.6/  29.9  & 303/  114&   19.9/   18.6 &  51  $\pm$    5 &     0.308  $\pm$   0.040  &  1  \\
  64&J132324.26+411515.01&   288/    111&    21.0/  17.6  &  340/  225&   21.9/   22.5 &  94  $\pm$    3 &     0.299  $\pm$   0.021  &  1  \\
  65&J132404.20+433407.14&    17/    198&    67.4/ 101.4  &  349/  186&   68.5/   88.2 & 130  $\pm$    6 &     0.155  $\pm$   0.018  &  1  \\
  66&J132713.87+285318.19&   285/    115&    55.8/  49.9  &   45/  225&   47.1/    4.0 &  90  $\pm$   16 &     0.113  $\pm$   0.032  &  1  \\
  67&J132939.95+181842.01&    52/    236&    23.1/  30.0  &  355/  160&   22.5/   19.8 &  $-$                  &   $-$  	     &  4  \\
  68&J133051.04+024843.10&   338/    158&    36.1/  44.1  &   16/  240&   27.5/   21.7 & 133  $\pm$    2   &     0.219  $\pm$   0.022&  1  \\
  69&J133636.06+431329.02&   270/     96&    28.1/  26.3  &  341/  140&   18.5/   29.2 &  $-$                  &   $-$  	     &  3  \\
  70&J134002.96+503539.72&    34/    211&    34.4/  33.2  &  276/   87&   18.2/   38.2 & 100  $\pm$    2 &     0.158  $\pm$   0.008  &  1  \\
  71&J134051.19+374911.74&   330/    1527&    36.7/  45.6  & 357/  180&   27.0/   30.4 &  43  $\pm$    2 &     0.306  $\pm$   0.009  &  1  \\
  72&J134353.97+193334.10&    69/    213&    25.6/  19.5  &  298/  119&    9.1/   24.2 &  $-$                  &   $-$  	     &  0$\dagger$  \\
  73&J135518.04+094022.90&   279/    105&    37.8/  35.6  &   62/  253&   23.5/   26.1 &  $-$                  &   $-$  	     &  0  \\
  74&J140349.79+495305.45&    28/    214&    40.7/  20.7  &  290/   64&   18.7/    9.5 &  $-$                  &   $-$  	     &  0  \\
  75&J140742.26+272207.66&   360/    203&    16.5/  15.5  &  292/  128&   16.8/   21.6 &  $-$                  &   $-$  	     &  0$^{\dagger}$ \\
  76&J141702.13+201903.30&    63/    230&    26.7/  20.7  &  309/  120&   24.4/    9.4 &  55  $\pm$    2 &     0.080 $\pm$    0.110  &  0  \\
  77&J142646.41+271223.63&   298/    118&    28.4/  27.2  &   62/  175&   23.5/   21.7 &  $-$                  &   $-$  	     &  0  \\
  78&J143756.45+351937.10&     2/    180&    19.9/  14.9  &  305/   54&   16.4/   17.0 &  $-$                  &   $-$  	     &  0  \\
  79&J144547.33$-$013045.77&  96/    266&    17.5/  26.0  &  315/  180&   17.3/   13.3 &  $-$                  &   $-$  	     &  0  \\
  80&J150016.24$-$045036.65&  86/    255&    29.3/  34.4  &   38/  218&   22.2/   21.6 &  $-$                  &   $-$  	     &  0$^{\dagger}$  \\
  81&J150636.54+074016.94&    42/    215&    22.4/  16.6  &  300/  135&   11.5/    4.5 &  $-$                  &   $-$  	     &  0  \\
  82&J150816.29+613756.32&    83/    262&    14.4/  13.8  &  346/  136&   14.8/   10.9 &  $-$                  &   $-$  	     &  0  \\
  83&J150855.22$-$073036.46&  14/    188&    33.4/  33.3  &   52/  225&   21.9/   18.5 &  $-$                  &   $-$  	     &  3$^{\dagger}$ \\
  84&J150904.13+212415.10&   270/     90&    26.0/  17.4  &   22/  198&   29.2/   20.2 & 140  $\pm$    2 &     0.227  $\pm$   0.014  &  1  \\
  85&J151149.30+045536.17&   315/    135&    18.6/  17.3  &   96/  251&   21.9/   18.9 &  $-$                  &   $-$  	     &  0  \\
  86&J151704.61+212242.14&   315/    135&    24.1/  19.0  &    1/  225&   16.9/   17.2 &  50  $\pm$    7 &     0.241  $\pm$   0.010  &  1  \\
  87&J152245.38$-$050404.36& 322/    142&    32.9/  31.6  &  287/  105&   25.6/   35.2 &  $-$                  &   $-$  	     &  4  \\
  88&J154202.85+121427.66&   303/    113&    18.7/  14.6  &   29/  246&   15.1/   11.8 &  $-$                  &   $-$  	     &  3 \\
  89&J154413.39+304401.16&    31/    210&    33.7/  35.1  &  356/  162&   27.4/   24.0 &  $-$                  &   $-$  	     &  0  \\
  90&J154719.43+213012.00&    85/    2266&    11.3/  13.8  & 336/  175&   11.8/   14.5 &  $-$                  &   $-$  	     &  3  \\
  91&J154842.66+014919.48&    34/    180&    42.2/  25.2  &   70/  222&   36.4/   31.0 &  $-$                  &   $-$  	     &  0  \\
  92&J155416.04+381132.57&    64/    238&    16.9/  25.9  &  360/  203&   13.3/   14.9 & 103  $\pm$    3 &     0.465  $\pm$   0.017  &  1  \\
  93&J160809.55+294514.92&   306/    126&    13.4/  15.3  &   41/  180&   20.0/   15.1 & 168  $\pm$    2 &     0.225  $\pm$   0.011  &  1  \\
  94&J160833.28+012231.04&    36/    2025&    16.9/  18.8  &  300/ 143&   10.9/   15.1 &  $-$               &   $-$  	             &  0  \\
  95&J162245.42+070714.69&   287/    107&    25.6/  25.9  &   72/  250&   24.1/   20.2 & 114  $\pm$    3 &     0.143  $\pm$   0.008  &  1  \\
  96&J164857.36+260441.26&   302/    135&    19.1/  23.2  &   65/  198&   12.6/   20.1 & 117  $\pm$    6 &     0.217  $\pm$   0.037  &  1  \\
  97&J171547.52+493840.22&   308/    131&    22.7/  20.8  &    5/  206&   14.8/   13.1 &  95  $\pm$    4 &     0.025 $\pm$    0.051  &  2$^{\dagger}$  \\
  98&J202855.27+003512.67&   283/    106&    41.9/  29.8  &   29/  212&   52.3/   34.2 &  58  $\pm$    6 &     0.228  $\pm$   0.023  &  1  \\
  99&J203459.54+005221.41&    59/    251&    22.4/  22.4  &  355/  180&   22.3/   13.1 &  88  $\pm$    5 &     0.136  $\pm$   0.030  &  1  \\
 100&J205823.53+031124.47&     4/    187&    32.0/  31.6  &   56/  232&   20.1/   22.0 &  $-$                  &   $-$  	     &  0  \\
 101&J210053.62$-$033516.66&  17/    200&    26.5/  19.4  &  328/   80&   18.6/   25.5 &  97  $\pm$    3 &     0.056 $\pm$    0.038  &  0$^{\dagger}$  \\
 102&J214731.06$-$035942.40& 311/     98&    17.5/  14.1  &  355/  204&   18.2/   12.1 &  68  $\pm$    5 &     0.114 $\pm$    0.082  &  0  \\
 103&J222802.33$-$065354.84& 272/     97&    24.6/  21.1  &   69/  249&   25.7/   25.8 &  $-$               &   $-$  	             &  0  \\
 104&J223628.89+042751.89&   358/    169&    51.6/  55.2  &   73/  245&   25.8/   21.9 & 145  $\pm$    8 &     0.245  $\pm$   0.034  &  1  \\
 105&J232020.30$-$075319.36&  33/    210&    42.2/  42.3  &  354/  166&   20.2/   23.5 &  73  $\pm$    3 &     0.430  $\pm$   0.004  &  1  \\
 106&J233259.28+024715.37&   342/    159&    21.3/  17.6  &   64/  234&   13.2/   16.6 &  70  $\pm$    2 &     0.065 $\pm$    0.047  &  0
\end{longtable*}    
}
\begin{minipage}{160mm}
{\small
  {$^a$Optical and radio position angles are measured from north to east.\\}
{$^b$Flag disription, 1= secure  detection ; 0 = faint source; 2= merging or two nearby objects; 3=  host galalxy not detected; 4= not covered in SDSS or DeCaLS; $\dagger$ optical position angle is measured from DECaLS images.}
}
\end{minipage}
\end{center}
\end{document}